\documentclass[global,twocolumn]{svjour}
\usepackage[T1]{fontenc}
\usepackage[latin9]{inputenc}
\usepackage{float}
\usepackage{amstext}
\usepackage{amssymb}
\usepackage{amsmath}
\usepackage{caption, ctable,threeparttable}
\usepackage{latexsym}
\usepackage{graphicx}
\begin{document}
\title{Dissipation, dephasing and quantum Darwinism in qubit systems with random unitary interactions}
\author{Nenad Balaneskovi$\textrm{\textrm{\ensuremath{\acute{\textrm{c}}}}}$\inst{1}
\thanks{\emph{email:} balaneskovic@gmx.net} 
Marc Mendler\inst{1}
%
}                     
%
%
\institute{\inst{1}Institut f\"ur Angewandte Physik, Technische Universit\"at Darmstadt, D-64289 Darmstadt, Germany 
}
\date{Received: date / Revised version: date}
%

\authorrunning{Nenad Balaneskovi$\textrm{\textrm{\ensuremath{\acute{\textrm{c}}}}}$ \emph{et. al.}}
\titlerunning{Random Unitary Operations, Dissipation, Dephasing and quantum Darwinism}
\maketitle
\abstract{
We investigate the influence of dissipation and decoherence on quantum Darwinism by generalizing Zurek's original qubit model of decoherence and the establishment of pointer states (Zurek, \cite{key-3}). Our model allows for repeated multiple qubit-qubit couplings between system and environment which are described by randomly applied two-qubit quantum operations inducing entanglement, dissipation and dephasing. The resulting stationary qubit states of system and environment are investigated. They exhibit the intricate influence of entanglement generation, dissipation and dephasing on this characteristic quantum phenomenon.
%
} 
\section{Introduction\label{A1}}

~~~Typically decoherence singles out interaction-robust states
of an open quantum system, the so called pointer states \cite{key-0}
that remain unaltered whenever an open quantum system is interacting with an environment.
By means of his concept of quantum Darwinism \cite{key-3} 
Zurek aims at exploring the corresponding characteristic features of quantum
information stored in the environment in a stable and redundant way
as the result of such a decoherence process. 

On the basis of simple qubit models in which each environmental qubit interacts with a
system qubit only once \cite{key-3} it
has been demonstrated that as a result of the quantum correlations established
between the open system and its environment
the stable pointer states of the open system become
strongly correlated with 
many-qubit copies of particular environmental states. 
Thus, the ``degree of objectivity'' of an open system's pointer
state can be quantified from the point of view of the environment
by simply counting the number of copies of this
information record deposited in the environmental
fragments.
High information redundancy of the system's pointer states within the
environment implies that some information about a subset of all system's quantum states containing the ``fittest'' states of the open system has been successfully distributed throughout all environmental fragments. Stated differently,
the environment stores redundant copies of information about these preferred
pointer states 
of the open system and thus accounts for their objective existence
\cite{key-0_1,key-0_2,key-0_3,key-0_4}).

Recently, this qubit based decoherence model has been generalized in order
to explore the question to which extent quantum Darwinism
depends on the fact that 
environmental qubits do not interact with each other and that each of them interacts with
a system qubit only once \cite{key-6}.
For this purpose the asymptotic dynamics of iterated random unitary quantum operations
involving unitary controlled-NOT operations
between randomly selected qubits has been investigated. In particular, these iterated quantum
operations allow for repeated multiple interactions between any qubit pairs.
Within this model it has been demonstrated
that characteristic features of quantum Darwinism are also observable in this more general
setting. However, it appears 
only for special classes of initial quantum states of systen and environment and
for mutually non interacting environmental qubits.

In view of Zurek's original work \cite{key-3} and this recent investigation \cite{key-6} 
the question arises, how the phenomenon of
quantum Darwinism is affected by more complicated quantum operations between system and environment and between environmental qubits which do not involve
unitary controlled-NOT operations alone.
It is a main purpose of this paper to address this issue.
For this purpose we investigate the influence of the
class of two-qubit quantum operations recently
proposed by
Scarani et al. \cite{key-4} 
which after successive applications
induce dissipation and dephasing 
of qubit network models \cite{key-4}.
We explore to which extent a repeated application of these quantum operations
between arbitrary qubit pairs influences
quantum Darwinism. 
In particular, it will be shown that iterative application of the dissipative
two-qubit operations of Scarani et al. suppresses
quantum Darwinism significantly. However, iterative application of the dephasing two-qubit operations
alone does not alter significantly the dynamics  in comparison with cases involving only iteratively applied unitary
controlled-NOT operations.

This paper is organized as follows:
In Sec. II 
Zurek's qubit toy model of quantum Darwinism is briefly summarized 
\cite{key-3}. In Sec. III we introduce the dissipative-dephasing
two-qubit operation previously motivated by Scarani et al. \cite{key-4} and embed it into the framework of the random unitary model.
In Sec. IV and V we compare quantum
Darwinism as obtainable from Zurek's qubit toy model (with dissipation and dephasing)
with the corresponding predictions of our modified random unitary model in the asymptotic long time limit of many iterations. 
In Sec. VI we give a brief summary of the most important results and an outlook on interesting future research problems.
Appendix \ref{AAG} presents analytic results of system-environment states
necessary for discussing quantum Darwinism within the framework of
Zurek's qubit toy model in the presence of dissipation and dephasing.
\section{Zurek's qubit toy model of quantum Darwinism\label{A2}}

~~~In this section we briefly summarize the basic ideas of the simplest
qubit toy model of quantum Darwinism (in the following referred to as "Zurek's (qubit) model"), as originally suggested by Zurek
\cite{key-3}, involving an open pure $k=1$-qubit system $S$ (with
$\left|\Psi_{S}^{\textrm{in}}\right\rangle =a\left|0\right\rangle +b\left|1\right\rangle $,
$\left(a,\, b\right)\in\mathbb{C}$, $\left|a\right|^{2}+\left|b\right|^{2}\overset{!}{=}1$ and $\overset{!}{=}$ meaning "should equal"),
which acts as a control-unit on its $\left(n\in\mathbb{N}\right)$-qubit
target (environment) $E\equiv\mathcal{E}_{1}\otimes\mathcal{E}_{2}\otimes...\otimes\mathcal{E}_{n}$.

Then, according to Zurek's qubit model the $S$-$E$ interaction has
to occur in the following manner (s. also \cite{key-6}):
\begin{enumerate}
\item Start with a pure $k=1$-qubit open $\hat{\rho}_{S}^{\textrm{in}}=\left|\Psi_{S}^{\textrm{in}}\right\rangle \left\langle \Psi_{S}^{\textrm{in}}\right|$
and an arbitrary $n$-qubit $\hat{\rho}_{E}^{\textrm{in}}$, where $\hat{\rho}_{SE}^{\textrm{in}}=\hat{\rho}_{S}^{\textrm{in}}\otimes\hat{\rho}_{E}^{\textrm{in}}$.
\item Apply the CNOT-gate $\hat{U}_{CNOT}\left|i\right\rangle _{S}\left|j\right\rangle _{E}=\left|i\right\rangle _{S}\left|i\oplus j\right\rangle _{E}$
(where $\oplus$ denotes addition modulo $2$), such that the $S$-qubit
$i$ interacts successively and only once with each qubit $j$ of
the environment $E$ until all $n$ $E$-qubits have interacted with
system $S$, resulting in an entangled state $\hat{\rho}_{SE}^{\textrm{out}}$.
\item Trace out successively (for example from \emph{right to left}) $\left(n-L\right)$
qubits in $\hat{\rho}_{E}^{\textrm{out}}$ and $\hat{\rho}_{SE}^{\textrm{out}}$ - this
yields the $L$-qubit $\hat{\rho}_{E_{L}}^{\textrm{out}}$ and $\hat{\rho}_{SE_{L}}^{\textrm{out}}$,
with $0<L\leq n$ and the $E$-fraction parameter $0<f=\frac{L}{n}\leq1$.
\item Compute the eigenvalue spectra $\left\{ \lambda_{1},...,\lambda_{d\left(f\right)}\right\} $
of $\hat{\rho}_{S}^{\textrm{out}}$, $\hat{\rho}_{E_{f}}^{\textrm{out}}$ and $\hat{\rho}_{SE_{f}}^{\textrm{out}}$
and the $f$-dependent von Neumann entropies \[H\left(\hat{\rho}\left(f\right)\right)=-\underset{i=1}{\overset{d\left(f\right)}{\sum}}\lambda_{i}\log_{2}\lambda_{i}\geq0,\,\underset{i=1}{\overset{d\left(f\right)}{\sum}}\lambda_{i}\overset{!}{=}1\]
(where $d\left(f\right)$ is the dimensionality of $\hat{\rho}\left(f\right)$
in question). 
\item Compute the normalized ratio $H\left(S:\, E_{f}\right)/H\left(S_{\textrm{class}}\right)$ depending on the $E$-fraction parameter $f$ and
involving the mutual information (MI) 
\begin{equation}
H\left(S:\, E_{f}\right)=H\left(S\right)+H\left(E_{f}\right)-H\left(S,\, E_{f}\right),\label{Gl 3.3}
\end{equation}
that quantifies the amount of the proliferated system's Shannon entropy
(>>classical information<<) \cite{key-0_2,key-3_1} 
\begin{equation}
H\left(S_{\textrm{class}}\right)=-\underset{i}{\sum}p_{i}\log_{2}p_{i}=H\left(\left\{ \left|\pi_{i}\right\rangle \right\} \right),\label{Gl 3.2}
\end{equation}
with probabilities $p_{i}=\mathbf{Tr}_{E}\left\langle \pi_{i}\right|\hat{\rho}_{SE}^{\textrm{class}}\left|\pi_{i}\right\rangle $ of an effectively decohered (>>quasi-classical<<)
system's $S$-state $\hat{\rho}_{S}^{\textrm{class}}$
emerging as partial traces with respect to the particular
$S$-pointer-basis $\left\{ \left|\pi_{i}\right\rangle \right\} $. In this context, $\hat{\rho}_{SE}^{\textrm{class}}$ is a special output state of the entire system, whose reduced system's density matrix $\hat{\rho}_{S}^{\textrm{class}}$ acquires a diagonal form (with vanishing outer-diagonal entries) in the limit $n\gg 1$ of effective decoherence (large environments) with respect to a specific computational basis. Compute the smallest vlue $f^{*}$ of the fraction parameter $f$ such that
\begin{equation}
\begin{array}{c}
R=1/f^{*}\,\left(0<f^{*}\leq1\right)\\
\textrm{with}\, H\left(S:\, E_{f=f^{*}}\right)\approx H\left(S_{\textrm{class}}\right)\left(n\gg1\right)
\end{array}\label{Gl 3.4}
\end{equation}
and $R$ denoting the redundancy of the measured $\left\{ \left|\pi_{i}\right\rangle \right\} $ in
the limit $n\gg1$ of effective decoherence. Finally, plot $H\left(S:\, E_{f}\right)/H\left(S_{\textrm{class}}\right)$ vs $0<f\leq1$
(Partial Information Plot (PIP) of MI).
\end{enumerate}

Now we look at the specific initial state \[\hat{\rho}_{SE}^{\textrm{in}}=\left|\Psi_{S}^{\textrm{in}}\right\rangle \left\langle \Psi_{S}^{\textrm{in}}\right|\otimes\left|0_{n}\right\rangle \left\langle 0_{n}\right|\]
with $\left|0_{n}\right\rangle \equiv\left|0\right\rangle ^{\otimes n}$
(ground state $\hat{\rho}_{E}^{\textrm{in}}=\left|0_{n}\right\rangle \left\langle 0_{n}\right|$)
\cite{key-3}. After allowing the one $S$-qubit to transform each
$E$-qubit via CNOT \emph{only once} until the entire environment
$E$ is affected, yielding $\forall L>0$

\begin{equation}
\label{Gl 3.5}
\left|\Psi_{SE_{L=n}}^{\textrm{out}}\right\rangle =a\left|0\right\rangle \otimes\left|0_{L=n}\right\rangle +b\left|1\right\rangle \otimes\left|1_{L=n}\right\rangle,
\end{equation}
with von Neumann-entropies
\begin{equation*}
\begin{array}{l}
H\left(S,\, E_{L}\right)=H\left(S_{\textrm{class}}\right)\cdot\left(1-\delta_{L,n}\right),\\
H\left(E_{L}\right)=H\left(S\right)=H\left(S_{\textrm{class}}\right)\,\forall L>0\\
H\left(S_{\textrm{class}}\right)=-\left|a\right|^{2}\log_{2}\left|a\right|^{2}-\left|b\right|^{2}\log_{2}\left|b\right|^{2}
\end{array}
\end{equation*}
and $p_{1}=\left|a\right|^{2},\,p_{2}=\left|b\right|^{2}$. (\ref{Gl 3.5}) demonstrates that $H\left(S:\, E_{f}\right)$, after
the $L$-th $E$-qubit has been taken into account ($f=L/n$), increases
from zero to \[H\left(S:\, E_{f}\right)\equiv H\left(S_{\textrm{class}}\right)\Rightarrow H\left(S:\, E_{f}\right)/H\left(S_{\textrm{class}}\right)=1,\]
implying that each fragment (qubit) of the environment $E$ supplies
complete information about the $S$-pointer observables $\left\{ \left|\pi_{i}\right\rangle \right\} \equiv\left\{ \left|\pi_{1}\right\rangle=\left|0\right\rangle ,\,\left|\pi_{2}\right\rangle=\left|1\right\rangle \right\} $.
Since the very first CNOT-operation forces the system $S$ to decohere
completely into its pointer-basis $\left\{\left|0\right\rangle,\,\left|1\right\rangle \right\} $,
one encounters the influence of quantum Darwinism on the observed
system $S$: from all possible $S$-states, which started their dynamics
within a pure $\hat{\rho}_{S}^{\textrm{in}}$, only diagonal elements of $\hat{\rho}_{S}^{\textrm{in}}$
survive constant monitoring of the environment $E$, whereas off-diagonal
elements of $\hat{\rho}_{S}^{\textrm{in}}$ vanish due to decoherence. Thus, the process of constant monitoring of system $S$ by its environment
$E$ selects a preferred (system's) pointer-basis $\left\{\left|0\right\rangle,\,\left|1\right\rangle \right\} $,
causing a continued increase of its redundancy $R$ throughout the
entire environment $E$. 

\begin{figure}[H]
\includegraphics[scale=0.054]{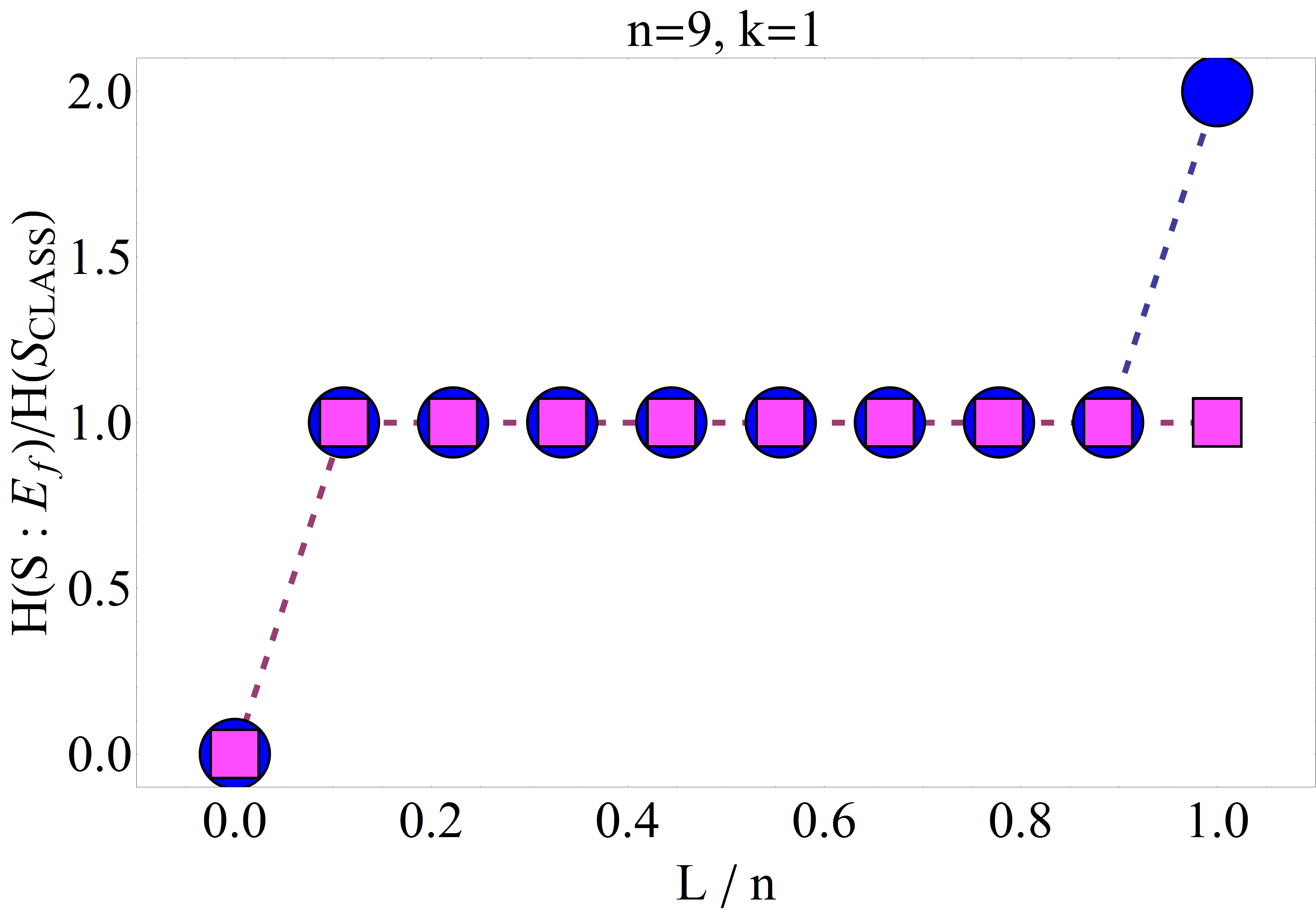}

\caption{Partial information plot (PIP) of mutual information (MI) and the redundancy $R$ of system's "classical" entropy  $H\left(S_{\textrm{class}}\right)$
stored in the $n$-qubit environment $E$ with respect to the $E$-fraction parameter $0<f=L/n\leq1$ after the $\widehat{U}_{\textrm{CNOT}}$-evolution
in accord with Zurek's ($\bullet$-dotted curve) \cite{key-3} and the random unitary
model ($\blacksquare$-dotted curve) \cite{key-6}: The initial $S$-$E$ state $\hat{\rho}_{SE}^{\textrm{in}}=\hat{\rho}_{S}^{\textrm{in}}\otimes\hat{\rho}_{E}^{\textrm{in}}$ involves a $k=1$
qubit pure $\hat{\rho}_{S}^{\textrm{in}}$ and $\hat{\rho}_{E}^{\textrm{in}}=\left|0_{n}\right\rangle \left\langle 0_{n}\right|$.
\label{Fig.1}}
\end{figure}

After decoherence we obtain \[H\left(S:\, E_{f}\right)=H\left(S_{\textrm{class}}\right)=H\left(E_{f}\right)=H\left(S,\, E_{f}\right),\]
valid for $0<f<1$, and the maximum \[H\left(S:\, E\right)=2H\left(S_{\textrm{class}}\right)\]
of MI (>>quantum peak<<), due to $H\left(S,\, E_{f=1}\right)=0$
after inclusion of the entire environment $E$ ($f=1$ ). Since each
$E$-qubit in Eq. (\ref{Gl 3.5}) is assumed to contain a perfect information
replica about the system's pointer-basis $\left\{\left|0\right\rangle,\,\left|1\right\rangle \right\} $, its
redundancy $R$ is given by the number of qubits in the environment
$E$, e.g. $R=n$. This constrains the form of MI in its PIP (see
Fig. \ref{Fig.1}, $\bullet$-dotted curve), which jumps from $0$
to $H\left(S_{\textrm{class}}\right)$ at $f=f^{*}=1/n$, continues
along the 'plateau' until $f=1-1/n$, before it eventually jumps up
again to $2H\left(S_{\textrm{class}}\right)$ at $f=1$.

Thus, $H\left(S:\, E_{f}\right)/H\left(S_{\textrm{class}}\right)\geq1$ indicates
high $R$ (objectivity) of $H\left(S_{\textrm{class}}\right)$ proliferated
throughout the environment $E$. Also, by intercepting already one
$E$-qubit we can reconstruct the system's pointer-basis $\left\{\left|0\right\rangle,\,\left|1\right\rangle \right\} $,
regardless of the order in which the $n$ $E$-qubits are being successively
traced out. Only if we need a small fraction of environment $E$ enclosing
maximally $n\cdot f^{*}=k\ll n$ $E$-qubits \cite{key-3}, to reconstruct
the system's pointer-basis $\left\{\left|0\right\rangle,\,\left|1\right\rangle \right\} $, quantum Darwinism
appears: apparently, it is not only important that the PIP-'plateau'
emerges, more relevant is its length $1/f^{*}$, corresponding to $R$
of the system's pointer-basis $\left\{\left|0\right\rangle,\,\left|1\right\rangle \right\} $.

The main question we aim to address in the following sections with respect to Zurek's and the random
unitary operations model (see section \ref{A3}) in the presence of dissipation and dephasing is:
are there input states $\hat{\rho}_{SE}^{\textrm{in}}$ that validate the relation
\begin{equation}
\frac{I\left(S:\, E_{f}\right)}{H\left(S_{class}\right)}=\frac{H\left(S\right)+H\left(E_{f}\right)-H\left(S,\, E_{f}\right)}{H\left(S_{class}\right)}\geq1,\label{Gl 3.5.2}
\end{equation}
with $H\left(S\right)\approx H\left(S_{class}\right)$ and $H\left(E_{f}\right)\geq H\left(S,E_{f}\right)$
at least for all $\left(k\leq L\leq [n\gg1]\right)$, regardless of
the order in which the $n$ $E$-qubits are being successively traced
out from the corresponding output state $\hat{\rho}_{SE}^{\textrm{out}}$? The inequality (\ref{Gl 3.5.2}) may be regarded as the main criterion for the occurrence of quantum Darwinism. Furthermore, in (\ref{Gl 3.5.2}) we also set, following Zurek's approach established in the course of his qubit toy model \cite{key-3} and without any loss of generality, the information deficit parameter $\delta$ to zero, since we are interested in a perfect information transfer between an open system of interest and its environment.
In order to answer this main question we discuss in the following the $f$-dependence of MI for different $\hat{\rho}_{SE}^{\textrm{in}}$ in the presence of dissipation and dephasing from the point of view
of Zurek's and the random unitary qubit model.

\section{Random Unitary Model of quantum Darwinism\label{A3}}

~~~In this section we will summarize the iterative random unitary model and generalize it to include, beside pure decoherence, dissipation and dephasing. 

When utilizing the decoherence mechanism in the course of an interaction between an open system and its environment, one usually thinks of the former as being monitored (observed) by the latter. In general, however, observations of an open quantum system of interest by its environment occur in nature in an uncontrolled way, generating dynamics which may even not be unitary at all. 

The easiest and most intuitive way to understand this uncontrolled monitoring of a system $S$ by its environment $E$ is to model the $S$-$E$ interaction of qubits constituting both subsystems as a Markov-chain of individual two-qubit "collisions". This Markov-chain approximation of quantum dynamics suggests that a realistic description of the decoherence process between open quantum systems can be achieved by assuming that mutual interactions between these quantum systems take place randomly, such that an output state of a quantum system under consideration depends only on its state immediately before the interaction and not on its entire evolution history.

As we shall see below, the iteratively applied random unitary evolution of open quantum systems will turn out to be the appropriate description of the Markov-chain approximation of quantum dynamics. This random unitary evolution will also generate non-unitary dynamics of an open system and its environment such that, as already demonstrated in \cite{key-6}, the quantum evolution emerging from Zurek's qubit model cannot be interpreted as the short-time limit of the random unitary evolution mechanism.

\begin{figure}[H]
\center\includegraphics[scale=0.34]{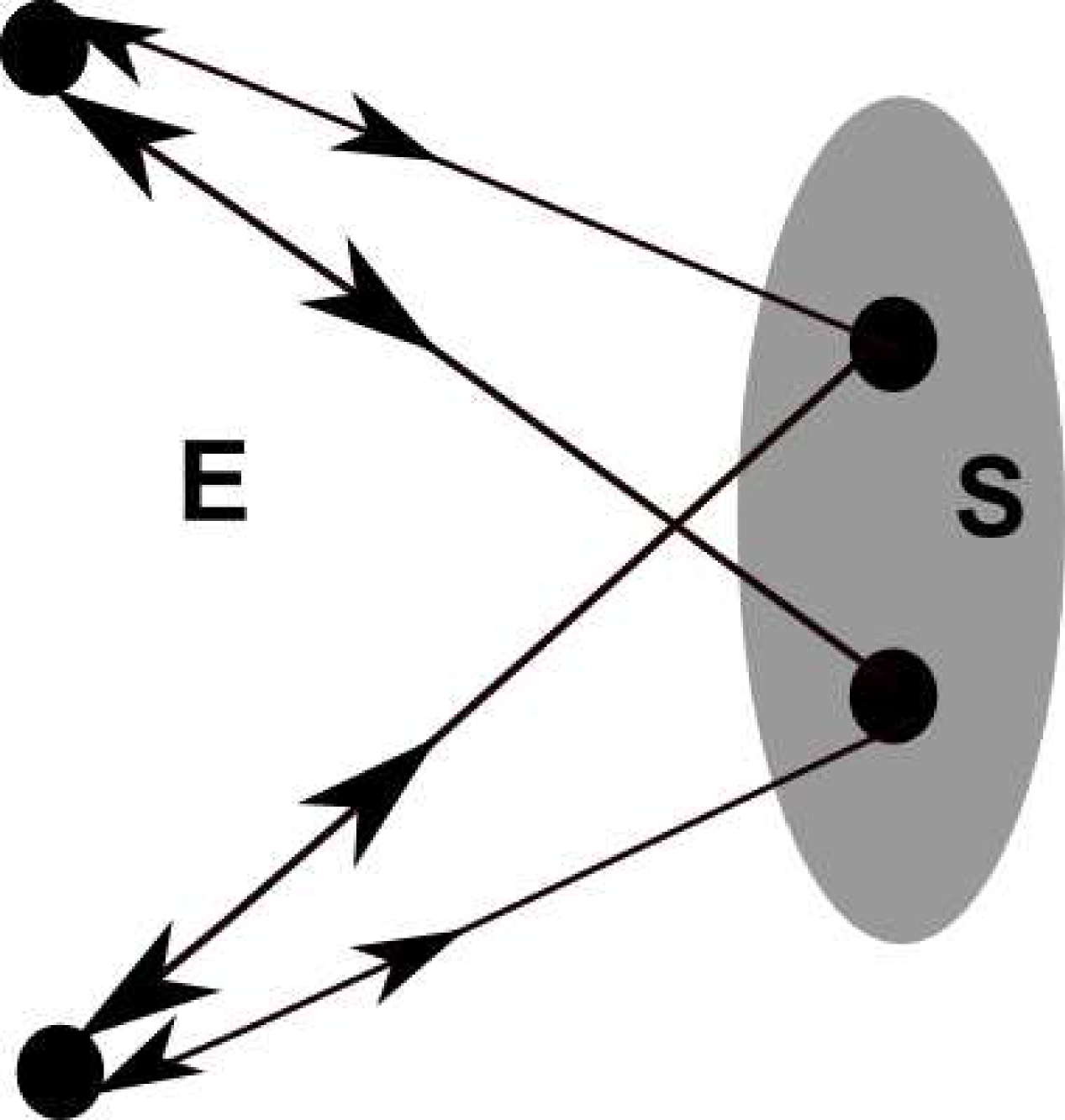}

\caption{Interaction digraph (ID) between system $S$ and environment $E$
including dissipation within the random unitary evolution formalism
\cite{key-2}. \label{Fig.2}}
\end{figure}

Namely, random unitary operations can be used to model the pure decoherence
of an open system $S$ with $k$ qubits (control, index $i$) interacting
with $n$ $E$-qubits (targets $j$) (as indicated in the directed
interaction graph (digraph) in Fig. \ref{Fig.2}) by the one-parameter
family of two-qubit \textquoteright{}controlled-U\textquoteright{}
unitary transformations \cite{key-6}. With respect to the standard one-qubit computational
basis $\left\{ \left|0\right\rangle,\,\left|1\right\rangle \right\} $ this $\phi$-parameter
family of two-qubit \textquoteright{}controlled-U\textquoteright{}
unitary transformations is given by
\begin{equation}
\widehat{U}_{ij}^{\left(\phi\right)}=\left|0\right\rangle _{i}\left\langle 0\right|\otimes\widehat{I}_{1}^{\left(j\right)}+\left|1\right\rangle _{i}\left\langle 1\right|\otimes\widehat{u}_{j}^{\left(\phi\right)},\label{Gl 2.1}
\end{equation}
where $\widehat{I}_{1}^{\left(j\right)}=\left|0\right\rangle _{j}\left\langle 0\right|+\left|1\right\rangle _{j}\left\langle 1\right|$ denotes the one-qubit identity matrix. Eq. (\ref{Gl 2.1}) indicates that only if an $S$-qubit $i$ should be in an excited
state, the corresponding targeted $E$-qubit $j$ hast to be modified
by a $\left(0\leq\phi\leq\pi\right)$-parameter gate $\widehat{u}_{j}^{\left(\phi\right)}$
\cite{key-2}, which for $\phi=\pi/2$ yields the CNOT-gate \cite{key-1,key-2,key-8}
\begin{equation}
\widehat{u}_{j}^{\left(\phi\right)}=\hat{\sigma}_{z}^{\left(j\right)}\cos
\phi+\hat{\sigma}_{x}^{\left(j\right)}\sin\phi\Rightarrow\widehat{u}_{j}^
{\left(\phi=\pi/2\right)}=\hat{\sigma}_{x}^{\left(j\right)},\label{Gl 2.2}
\end{equation}
with Pauli matrices $\hat{\sigma}_{l}$, $l\in\left\{ x,\, y,\, z\right\} $.

Arrows of
the interaction digraph (ID) in Fig. \ref{Fig.2} from $S$- to $E$-qubits
represent two-qubit interactions $\widehat{u}_{j}^{\left(\phi\right)}$
between randomly chosen qubits $i$ (of system $S$) and $j$ (of environment $E$) with initial probability distribution
$\{p_{e}\}$ (where $1>p_{e}>0$ $\forall e$) used to weight the edges $e=(ij)\in M$ of the digraph, with $M$ being a set of all edges in the ID. All
interactions are well separated in time. The $S$-qubits do not interact
among themselves. Furthermore, in this paper we also assume that the
$E$-qubits, as in Zurek's model, are not allowed to interact among
themselves. In other words, $e$ ranges over pairs $(i,\,j)$ of qubits with an index $i$ labelling a system qubit and an index $j$ labelling an environmental qubit.

Arrows of the interaction digraph (ID) in Fig. \ref{Fig.2}
from $E$- to $S$-qubits denote, on the other hand, a dissipative-dephasing feedback of the environment $E$ with respect to the system $S$ represented
by two qubit interactions \cite{key-4}
\begin{equation}
\label{Gl 2.3}
\begin{array}{l}
\hat{U}_{ij}^{\textrm{Diss}}\left(\alpha_{1},\,\alpha_{2},\,\gamma\right)=\exp\left[\frac{i}{2}\left(\hat{H}_{\gamma}^{\alpha_{1},\,\alpha_{2}}\right)\right],
\end{array}
\end{equation}
with real-valued dissipation strengths $0\leq\alpha_{1}+\alpha_{2}\leq\pi$,
a dephasing rate $0\leq\gamma\leq\pi$ and the Hamilton operator given by $\hat{H}_{\gamma}^{\alpha_{1},\,\alpha_{2}}=\alpha_{1}\hat{\sigma}_{x}^{\left(i\right)}\otimes\hat{\sigma}_{x}^{\left(j\right)}+
\alpha_{2}\hat{\sigma}_{y}^{\left(i\right)}\otimes\hat{\sigma}_{y}^{\left(j\right)}-\gamma
\hat{\sigma}_{z}^{\left(i\right)}\otimes\hat{\sigma}_{z}^{\left(j\right)}$.

The
total unitary two-qubit operation, accounting also for the dissipative-dephasing
effects in the course of the random unitary evolution, is then given
by
\begin{equation}
\hat{U}_{ij}^{\textrm{Tot}}\left(\alpha_{1},\,\alpha_{2},\,\gamma\right)=\hat{U}_{ij}^{\left(\phi=\pi/2\right)}\hat{U}_{ij}^{\textrm{Diss}}\left(
\alpha_{1},\,\alpha_{2},\,\gamma\right).\label{Gl 2.4}
\end{equation}
Thus, the randomly applied unitary operations of Eq. (\ref{Gl 2.4}) model the interplay between decoherence,
dissipation and dephasing with respect to an open $k$-qubit system $S$ and its
$n$-qubit environment $E$.

 In order to model the realistic measurement
process of system $S$ by its environment $E$ we let an initial state
$\hat{\rho}_{SE}^{\textrm{in}}$ evolve by virtue of the following iteratively
applied random unitary quantum operation (completely positive unital map) of the form $\mathcal{P}_{\textrm{Tot}}\left(\centerdot\right)\equiv\sum_{e\in M}K_{e}^{\textrm{Tot}}\left(\centerdot\right)K_{e}^{\textrm{Tot}\dagger}$, with Kraus-operators given by the relation \\ $K_{e}^{Tot}:=\sqrt{p_{e}}\hat{U}_{e}^{\textrm{Tot}}\left(\alpha_{1},\,\alpha_{2},\,\gamma\right)$
\cite{key-1,key-2,key-8}:

1. The quantum state $\hat{\rho}(N)$ after $N$ iterations is changed
by the $(N+1)$-th iteration to the quantum state (quantum Markov
chain)
\begin{equation}
\label{Gl 2.5}
\widehat{\rho}\left(N+1\right)=\sum_{e\in M}p_{e}\hat{U}_{e}^{\textrm{Tot}}\widehat{\rho}\left(N\right)\hat{U}_{e}^{\textrm{Tot}}
\equiv\mathcal{P}_{\textrm{Tot}}\left(\widehat{\rho}\left(N\right)\right).
\end{equation}

2. In the asymptotic limit $N\gg1$ $\widehat{\rho}\left(N\right)$
is independent of the initial probability distribution $\{p_{e},\, e\in M\}$, with $1>p_{e}>0$ $\forall e$, and (for $N\rightarrow\infty$) determined by
linear attractor spaces with mutually orthonormal solutions (index
$i$) $\hat{X}_{\lambda,i}$ of the eigenvalue equation
\begin{equation}
\hat{U}_{e}^{\textrm{Tot}}\left(\alpha_{1},\,\alpha_{2},\,\gamma\right)\widehat{X}_{\lambda,i}\hat{U}_{e}^{\textrm{Tot}}\left(\alpha_{1},\,\alpha_{2},\,\gamma\right)=\lambda\widehat{X}_{\lambda,i},\,\forall\, e\in M,\label{Gl 2.6}
\end{equation}
as subspaces of the total $S$-$E$-Hilbert space $\mathcal{H}_{SE}=\mathcal{H}_{S}\otimes\mathcal{H}_{E}$
to the eigenvalues $\lambda$ (with $\left|\lambda\right|=1$) \cite{key-1,key-8}.

3. For known attractor spaces $\mathcal{A}_{\lambda}$ we get from
an initial state $\hat{\rho}_{SE}^{\textrm{in}}$ the resulting $S$-$E$-state
$\widehat{\rho}_{SE}^{\textrm{out}}=\widehat{\rho}_{SE}\left(N\gg1\right)$
spanned by $\widehat{X}_{\lambda,i}$
\begin{equation}
\widehat{\rho}_{SE}^{\textrm{out}}=\mathcal{P}^{N}_{\textrm{Tot}}\left(\widehat{\rho}_{SE}^{\textrm{in}}
\right)=\underset{\left|\lambda\right|=1,\, i=1}{\overset{d^{\lambda}}{\sum}}\lambda^{N}\textrm{Tr}\left\{ \widehat{\rho}_{SE}^{\textrm{in}}\widehat{X}_{\lambda,i}^{\dagger}\right\} \widehat{X}_{\lambda,i},\label{Gl 2.7}
\end{equation}
where $d^{\lambda}$ denotes the dimensionality of the attractor space
$\mathcal{A}_{\lambda}$ with respect to the eigenvalue $\lambda$. The entire attractor space $\mathcal{A}$ is then given by $\mathcal{A}=\underset{\lambda}{\oplus}\mathcal{A}_{\lambda}.$

Finally, some further remarks about the above random unitary evolution algorithm are in order before we turn our attention in the forthcoming section \ref{A4} to the impact of symmetric dissipation on quantum Darwinism in the course of Zurek's and the random unitary evolution model: 
\begin{itemize}
\item The attractor space $\mathcal{A}$ is a subspace of the total $S$-$E$-Hilbert space $\mathcal{H}_{SE}=\mathcal{H}_{S}\otimes\mathcal{H}_{E}$ containing itself mutually orthogonal subspaces determined from Eq. (\ref{Gl 2.6}) with respect to eigenvalues $\lambda$ and spanned by the corresponding attractor states $\widehat{X}_{\lambda,i}$. Each subspace $\mathcal{A}_{\lambda}$ of $\mathcal{A}$ is thus a bassain of attraction for the dynamics of a certain inital $S$-$E$ state in the asymptotic limit of many system-environment interactions, such that $\mathcal{A}$ can be expressed as a direct sum over all $\mathcal{A}_{\lambda}$. In other words, in the asymptotic limit of many interactions the dynamics of an initial $S$-$E$ state subject to the random unitary evolution remains confined ("captured") within the attractor space $\mathcal{A}$ such that its corresponding final state can be decomposed by means of attractor states $\widehat{X}_{\lambda,i}$ in accord with Eq. (\ref{Gl 2.7}).
\item When talking about the asymtotic limit of many iterations we need to distinguish between the limit $N\gg1$, reserved for numerical computation according to Eq. (\ref{Gl 2.5}), and the limit $N\rightarrow\infty$, used when referring to analytic (attractor space) solutions from Eq. (\ref{Gl 2.6}). Nevertheless, for the sake of consistency $\widehat{\rho}_{SE}^{\textrm{out}}$, obtained from the random unitary evolution in Eq. (\ref{Gl 2.5}) with a sufficiently high $N$-value, has to converge to analytic results emerging from Eq. (\ref{Gl 2.6}) and (\ref{Gl 2.7}) in the (strict) limit $N\rightarrow\infty$. Whenever necessary, we will explicitly point out this subtlety regarding the expression "the asymptotic limit" in the forthcoming sections.
\end{itemize} 
%

\section{Random unitary operations perspective on quantum Darwinism: pure
decoherence vs symmetric dissipation ($\alpha_{1}=\alpha_{2}=\alpha\geq\gamma=0$)\label{A4}}

~~~In this section we discuss the impact of symmetric dissipation (Eq. (\ref{Gl 2.4}) with $\alpha_{1}=\alpha_{2}=\alpha>\gamma=0$) on the
$f$-dependence of MI in the framework of the random unitary evolution
model in the asymptotic limit of many iterations of Eq. (\ref{Gl 2.5}) ($N\gg1$)
and compare it with conclusions that can be drawn from Zurek's qubit-model
of quantum Darwinism. We will see that quantum Darwinism disappears in the framework of the random unitary model as soon as $\alpha>0$. 

To be more specific, we repeat the main question we aim
to address with respect to Zurek's and the random unitary operations
model: do initial states $\hat{\rho}_{SE}^{\textrm{in}}$ exist that, despite
of dissipation, validate the quantum Darwinistic relation \cite{key-6}

\begin{equation}
\frac{H\left(S:\, E_{f}\right)}{H\left(S_{class}\right)}=\frac{H\left(S\right)+H\left(E_{f}\right)-H\left(S,\, E_{f}\right)}{H\left(S_{\textrm{class}}\right)}\geq1,\label{Gl 3.5.1}
\end{equation}
with $H\left(S\right)\approx H\left(S_{\textrm{class}}\right)$ and $H\left(E_{f}\right)\geq H\left(S,\, E_{f}\right)$,
at least $\forall\,\left(k\leq L\leq n\gg1\right)$, regardless of
the order in which the $n$ $E$-qubits are being successively traced
out from the resulting final state $\hat{\rho}_{SE}^{\textrm{out}}$? 

Let us first start with pure decoherence ($\alpha=0$). Within the
random unitary model we are led to another type of PIP-behavior:
inserting $\hat{\rho}_{SE}^{\textrm{in}}$ from Fig. \ref{Fig.1} into Eq. (\ref{Gl 2.5})
we obtain for $\alpha=0$ (with, for instance, $\left|a\right|^{2}=\left|b\right|^{2}=1/2$
and $p_{e}=1/\left|M\right|$ $\forall e$) after $N\gg1$ iterations the PIP in
Fig. \ref{Fig.1} ($\blacksquare$-dotted curve), which suggests that
Zurek\textquoteright{}s MI-\textquoteright{}plateau\textquoteright{}
\cite{key-3} appears only in the limit $N\rightarrow\infty$ \cite{key-6}. 

\begin{figure}[H]
\center\includegraphics[scale=0.051]{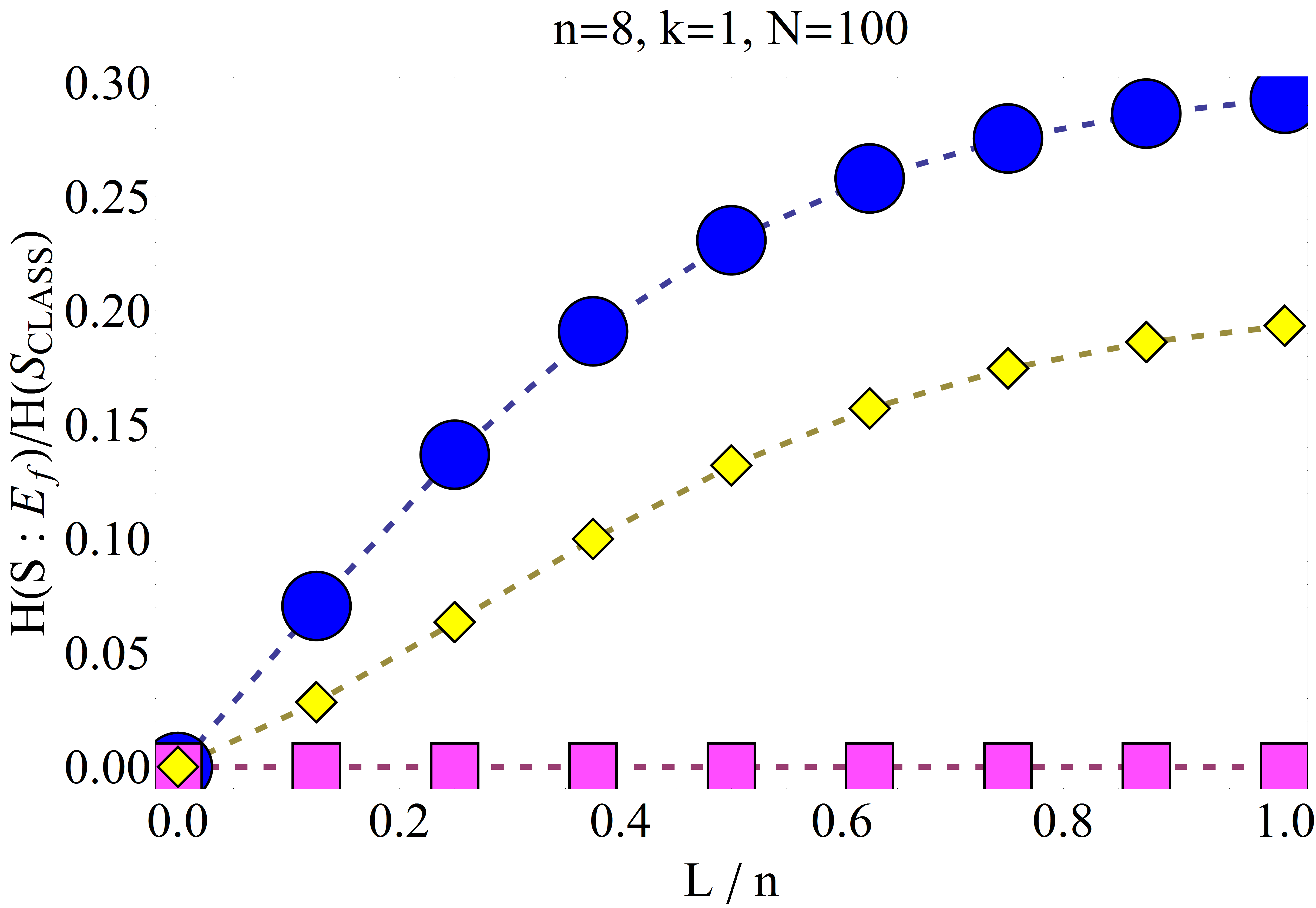}

\caption{Random unitary evolution model: PIP of MI vs $0<f\leq1$ for $\alpha=\pi/2$,
$\hat{\rho}_{SE}^{\textrm{in}}=\hat{\rho}_{S}^{\textrm{in}}\otimes\hat{\rho}_{E}^{\textrm{in}}$,
with a $k=1$ pure $\hat{\rho}_{S}^{\textrm{in}}$ and $\hat{\rho}_{E}^{\textrm{in}}=\left|0_{n}\right\rangle \left\langle 0_{n}\right|$
($\bullet$-dotted curve), $\hat{\rho}_{E}^{\textrm{in}}=0.5\cdot\left(\left|0_{n}\right\rangle \left\langle 0_{n}\right|+\left|1_{n}\right\rangle \left\langle 1_{n}\right|\right)$
($\blacklozenge$-dotted curve), $\hat{\rho}_{E}^{\textrm{in}}=2^{-n}\cdot\hat{I}_{n}$
($\blacksquare$-dotted curve). \label{Fig.3}}
\end{figure}

\begin{figure}[H]
\center\includegraphics[scale=0.045]{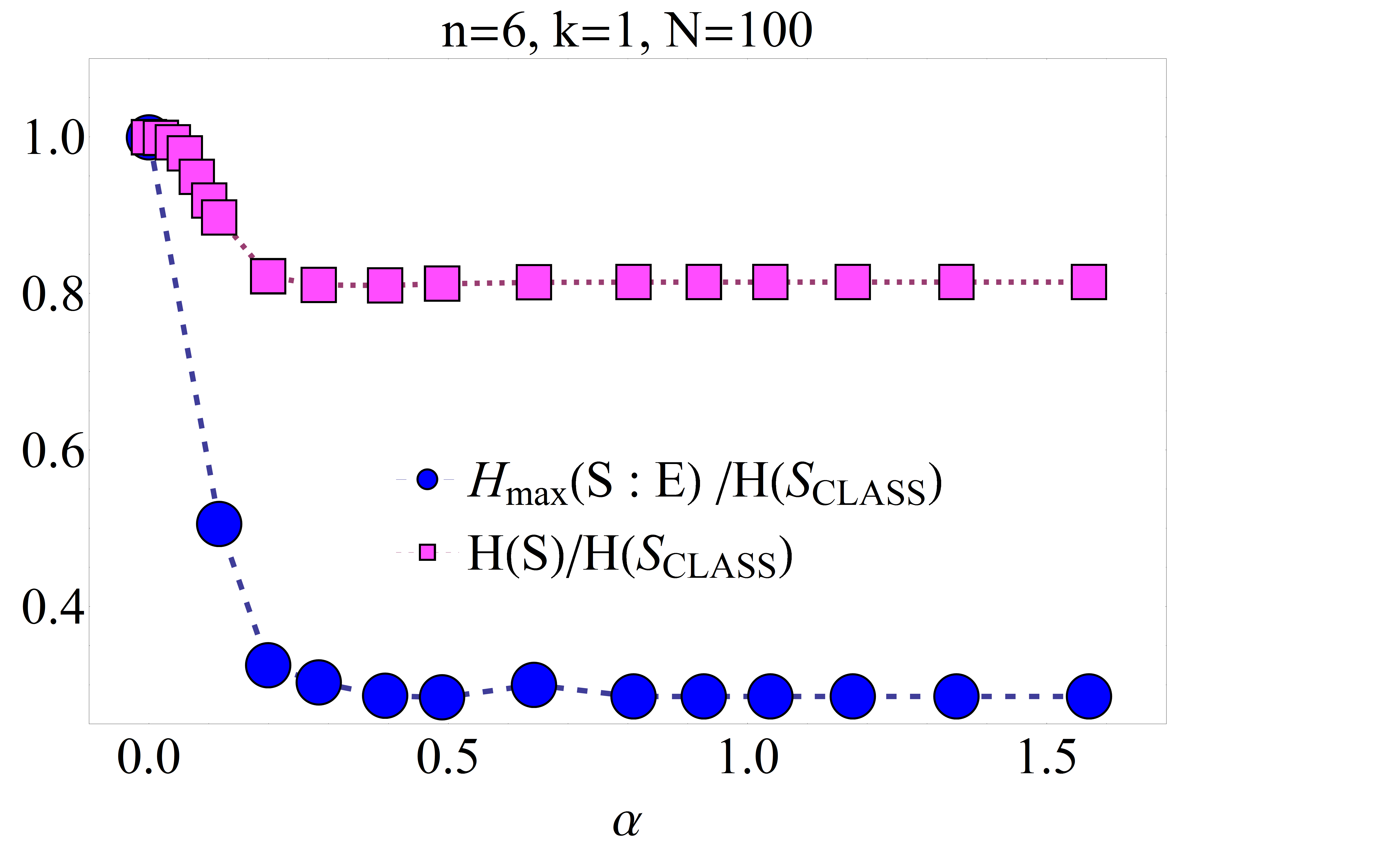}

\caption{Random unitary evolution model: maximal value $H\left(S:\, E_{L=n}\right)=H_{max}\left(S:\, E\right)$
of MI and $H\left(S\right)$ from Fig. \ref{Fig.3} vs $0<\alpha\leq\pi/2$ for $L=n=6$. \label{Fig.4}}
\end{figure}

For $\alpha>0$, $\left|a\right|^{2}=\left|b\right|^{2}=1/2$ and
$p_{e}=1/\left|M\right|$ $\forall e$ we obtain, applying to $\hat{\rho}_{SE}^{\textrm{in}}$
from Fig. \ref{Fig.1} iteratively Eq. (\ref{Gl 2.5}) $N\gg1$ times,
the PIP in Fig. \ref{Fig.3} ($\bullet$-dotted curve),
which suggests that for $N\gg1$, $\alpha>0$, $L=n\geq5$ one has
\[
\begin{array}{l}
\underset{N\gg1}{\lim}H\left(S:\, E_{L=n}\right)/H\left(S_{\textrm{class}}\right)\approx 0.3\\
\underset{\alpha\rightarrow\pi/2}{\lim}H\left(S\right)/H\left(S_{\textrm{class}}\right)\approx 0.8\\
\underset{N,\, n\gg1}{\lim}H\left(S\right)<H\left(S_{\textrm{class}}\right)=1.
\end{array}
\]
 Thus, for $\pi/2\geq\alpha>0$ there is, even for Zurek's quantum
Darwinism-compliant initial state $\hat{\rho}_{SE}^{\textrm{in}}$ associated with Eq. (\ref{Gl 3.5}),
which leads to the MI-plateau in Fig. \ref{Fig.1} ($\blacksquare$-dotted
curve) in the limit $N\rightarrow\infty$ (see \cite{key-6}), no quantum
Darwinism within the random unitary evolution model.
This also follows from the $\alpha$-behavior of $H\left(S:\, E_{L=n}\right)$
for a fixed $L=n=6$ and $N=100\gg1$ in Fig. \ref{Fig.4} ($\bullet$-dotted
curve): as long as $\alpha>0$, $H\left(S:\, E_{L=n}\right)$ will
fall below $H\left(S_{\textrm{class}}\right)=1$ (Eq. (\ref{Gl 2.4})
with $\alpha>0$ is an 'imperfect' copy-machine). Furthermore, the
corresponding $H\left(S\right)$ in Fig. \ref{Fig.4} ($\blacksquare$-dotted
curve) falls for $\alpha>0$ below $H\left(S_{\textrm{class}}\right)=1$, indicating
loss of information about $\left\{ \left|\pi_{i}\right\rangle \right\} $
of system $S$. 

\begin{figure}[H]
\center\includegraphics[scale=0.050]{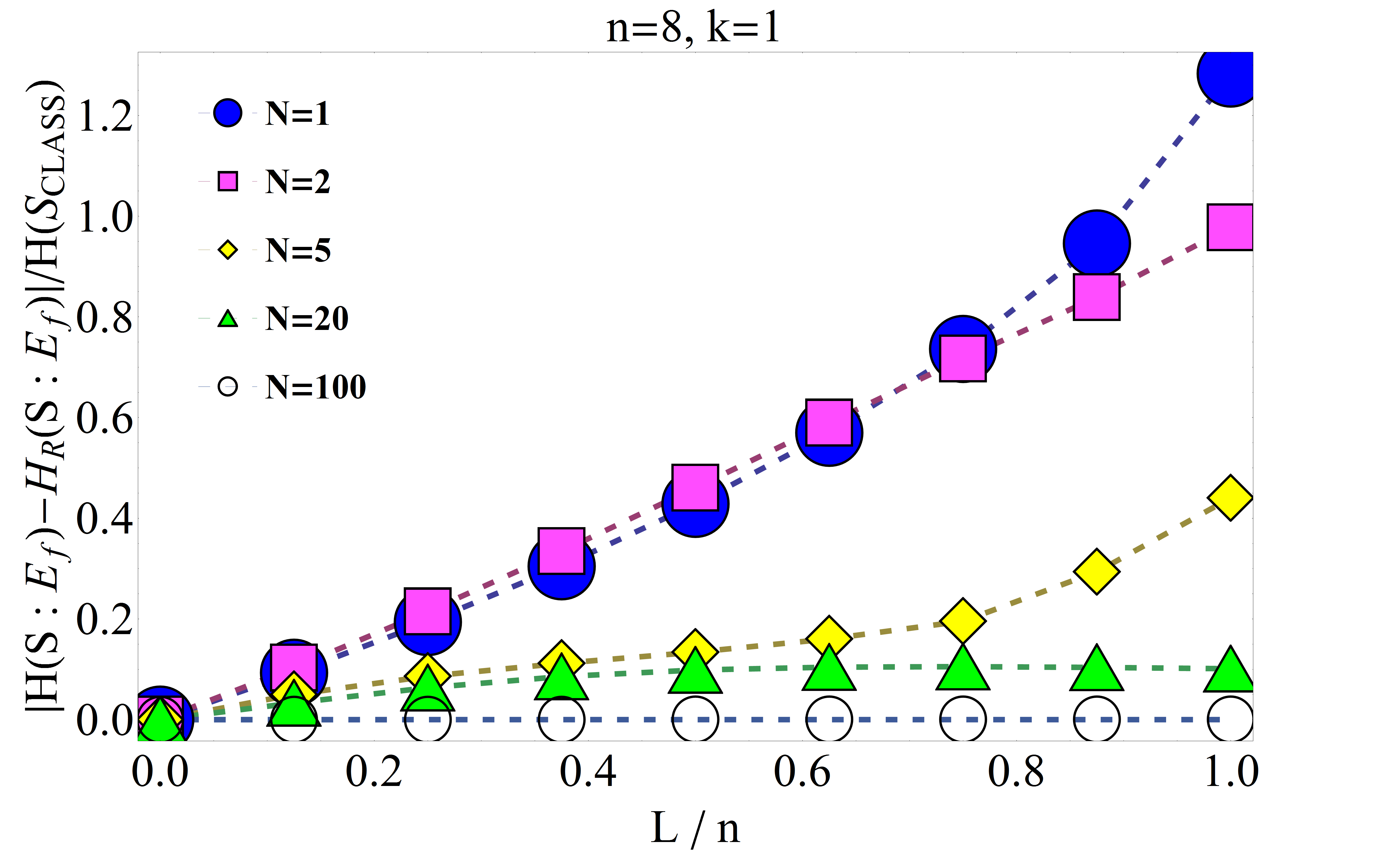}
\caption{PIP (for $\alpha=\pi/2$) of the MI-difference $\left|H\left(S:\, E_{L}\right)-H_{R}\left(S:\, E_{L}\right)\right|/H\left(S_{\textrm{class}}\right)$
with respect to the random unitary evolution of $\hat{\rho}_{SE}^{\textrm{in}}$
from Fig. \ref{Fig.1}, leading (with Eq. (\ref{Gl 2.4})) to $H\left(S:\, E_{L}\right)$
and (with $\hat{U}_{ij}^{\textrm{Diss}}
\left(\alpha\right)\hat{U}_{ij}^{\left(\phi=\pi/2\right)}$)
to $H_{R}\left(S:\, E_{L}\right)$. \label{Fig.5}}
\end{figure}

Also, exchanging in Eq. (\ref{Gl 2.4}) the order of application of Eq. (\ref{Gl 2.1})-(\ref{Gl 2.2})
and Eq. (\ref{Gl 2.3}) does not affect the behavior of $H\left(S:\, E_{L}\right)/H\left(S_{\textrm{class}}\right)$
with respect to $f$ for $N\gg1$: namely, in the limit $N\gg1$ the
PIP of Fig. \ref{Fig.5}, which displays the difference between the
PIP of $H\left(S:\, E_{L}\right)$ from Fig. \ref{Fig.3} and the PIP
of $H_{R}\left(S:\, E_{L}\right)$ (emerging from the random unitary
evolution of $\hat{\rho}_{SE}^{\textrm{in}}$ associated with Eq. (\ref{Gl 3.5})
with respect to the reversed operator
order $\hat{U}_{ij}^{\textrm{Diss}}\left(\alpha
\right)\hat{U}_{ij}^{\left(\phi=\pi/2\right)}$),
tends to zero.

However, applying Eq. (\ref{Gl 2.4}) and $\hat{U}_{ij}^{\textrm{Diss}}\left(\alpha=\pi/2\right)\hat{U}_{ij}^{\left(\phi=\pi/2
\right)}$
to $\hat{\rho}_{SE}^{\textrm{in}}$ associated with Eq. (\ref{Gl 3.5}) in accord with Zurek's
qubit model of quantum Darwinism we obtain

\begin{equation}
\begin{array}{c}
\hat{U}_{ij}^{\left(\phi=\pi/2\right)}\hat{U}_{ij}^{\textrm{Diss}}\left(\alpha=\pi/2\right)\left(\left|\Psi_{S}^{\textrm{in}}\right\rangle \otimes\left|0_{n}\right\rangle \right)=\\
=a\left|0\right\rangle \left|0_{n}\right\rangle +i\cdot b\left|0\right\rangle \left|10_{n-1}\right\rangle \Rightarrow\\
\\
H\left(S,\,E_{L}\right)=H\left(E_{L}\right)=H\left(S\right)=0\,\forall\,\left(0<L\leq n\right)\\
\\
\hat{U}_{ij}^{\textrm{Diss}}\left(\alpha=\pi/2\right)\hat{U}_{ij}^{\left(\phi=\pi/2\right)}\left(\left|\Psi_{S}^{\textrm{in}}\right\rangle \otimes\left|0_{n}\right\rangle \right)=\\
=a\left|0\right\rangle \otimes\left|0_{L=n}\right\rangle +b\left|1\right\rangle \otimes\left|1_{L=n}\right\rangle .
\end{array}\label{Gl 4.1}
\end{equation}
(\ref{Gl 4.1}) shows that the order of the dissipative and the CNOT-operator in
Eq. (\ref{Gl 2.4}) does matter within Zurek's model of quantum Darwinism.

Now we turn to other $\hat{\rho}_{SE}^{\textrm{in}}$ and their PIPs in Fig.
\ref{Fig.3} that we let evolve as in Eq. (\ref{Gl 2.1})-(\ref{Gl 2.5})
for $\alpha>0$, $\left|a\right|^{2}=\left|b\right|^{2}=1/2$ and
$p_{e}=1/\left|M\right|$ $\forall e$. 
The $\blacklozenge$-dotted curve in Fig.\ref{Fig.3} demonstrates that introducing correlations into $\hat{\rho}_{E}^{\textrm{in}}$
suppresses quantum Darwinism within the random unitary model, as for
$\alpha=0$ in \cite{key-6}: since 
\[
\underset{N\gg1}{\lim}H\left(S:\, E_{L=n}\right)/H\left(S_{\textrm{class}}\right)=0.2
\]
for a fixed $\alpha>0$ and $L=n\geq5$, no quantum Darwinism appears.

This becomes apparent if we look at MI $H\left(S:\, E_{L}\right)/$\\$H\left(S_{\textrm{class}}\right)$
from the $\blacksquare$-dotted curve in Fig. \ref{Fig.3} (with initial $E$-state $\hat{\rho}_{E}^{\textrm{in}}=2^{-n}\cdot\hat{I}_{n}=2^{-n}\hat{I}_{1}^{\otimes n}$): one has in this case 
\[
\underset{N\gg1}{\lim}H\left(S:\, E_{L=n}\right)/H\left(S_{\textrm{class}}\right)=0
\]
 for a fixed $\alpha>0$ and $L=n\geq5$, i.e. completely mixed $\hat{\rho}_{E}^{\textrm{in}}$
suppress quantum Darwinism (no MI-\textquoteright{}plateau\textquoteright{}),
as in \cite{key-6}.

Interestingly, for an entangled initial state (with an $S$-probability distribution $\left|a\right|^{2}=\left|b\right|^{2}=1/2$)
\begin{equation}
\left|\Psi_{SE}^{\textrm{in}}\right\rangle =a\left|0_{k=1}\right\rangle \left|s_{1}^{L=n}\right\rangle +b\left|1_{k=1}\right\rangle \left|s_{2}^{L=n}\right\rangle \label{Gl 4.1.1_ent}
\end{equation}

\noindent involving CNOT-invariant $E$-states
\[
\left|s_{m}^{n}\right\rangle =\left|s_{m}\right\rangle ^{\otimes n}=\left(\sqrt{2}\right)^{-n}\left(\left|0\right\rangle + (-1)^{m+1}\left|1\right\rangle \right)^{\otimes n},
\] 
with $\hat{\sigma}_{x}\left|s_{m}\right\rangle =\left(-1\right)^{m+1}\left|s_{m}\right\rangle$, $m\in\left\{ 1,\,2\right\} $
and $\left\langle s_{1}|s_{2}\right\rangle =0$,
we obtain $H\left(S:\, E_{L}\right)/$ $H\left(S_{\textrm{class}}\right)$
that leads to quantum Darwinism if $\alpha=0$ (as in Fig. \ref{Fig.1})
\cite{key-6}, but behaves in the limit $N\gg1$ for $n\geq5$ and
a fixed $\alpha=\pi/2$ with respect to $f=L/n$ as the $\blacksquare$-dotted
curve in Fig. \ref{Fig.3}.

Finally, for $\hat{\rho}_{SE}^{\textrm{in}}$ from Fig. \ref{Fig.1} with $N=100$,
$k=2,\,3$, $p_{i}=2^{-k}\,\forall\, i\in\left\{ 0,\,...,\,2^{k}-1\right\} $
and $H\left(S_{\textrm{class}}\right)=k$ (see (\ref{Gl 3.2}) above) we obtain from the random unitary
evolution the PIP as in Fig. \ref{Fig.6}, showing that for $L=n\geq5$ and $N\gg 1$
\[
\underset{(k\sim n)\gg1}{\lim}H\left(S:\, E_{L=n}\right)/H\left(S_{\textrm{class}}\right)=2^{-k},
\]
i.e. no quantum Darwinism $\forall k>1$ (as in \cite{key-6} with
$\alpha=0$).

\textbf{Summary}

Results in Fig. \ref{Fig.3}-\ref{Fig.6} and Eq. (\ref{Gl 4.1}) indicate that in case of non-vanishing, symmetric dissipation (without dephasing) Zurek's and the random unitary evolution model lead to different results when it comes to the appearance of quantum Darwinism: whereas in the course of Zurek's evolution qubit model the order in which pure decoherence and dissipation act upon a given $S$-$E$ inital state matters (i.e. if one applies first the CNOT-operation and then the dissipation operation to $\hat{\rho}_{SE}^{\textrm{in}}$, quantum Darwinism appears, otherwise no MI-'plateau' appears), quantum Darwinism would never appear in the asymptotic limit of the random unitary evolution, regardless of the order in which CNOT and dissipation act upon $\hat{\rho}_{SE}^{\textrm{in}}$. We will confirm these results in the forthcoming subsection by numerically reconstructing the symmetric dissipative attractor space.

\subsection{Numerical reconstruction of the symmetric dissipative attractor space\label{A4.1}}

~~~Now we reconstruct the symmetric dissipative attractor space of the random
unitary model by means of numerical simulations. Therefore, we start
with (\ref{Gl 2.4}) that has (for $\alpha>0$) four eigenvalues

$\lambda\in\left\{ 1,\,-1,\,0.5\cdot\left(1+\cos\alpha\pm i\sqrt{4-\left(1+\cos\alpha\right)^{2}}\right)\right\} $.
When looking at the matrix structure of $\hat{\rho}_{SE}^{\textrm{in}}$ from
Fig. \ref{Fig.1} after $N\gg1$ random unitary iterations of Eq. (\ref{Gl 2.5})
we notice that the outer-diagonal entries of $\hat{\rho}_{SE}^{\textrm{out}}$
from Fig. \ref{Fig.3} ($\bullet$-dotted curve) tend to zero $\sim n^{-N}\,\forall\alpha>0$,
such that we obtain
\begin{equation}
\begin{array}{l}
\hat{\rho}_{SE}^{\textrm{out}}\left(N\gg1\right)=\frac{\left|b\right|^{2}}{2^{n+1}-1}\cdot\hat{I}_{1+n}\\
+\left(\left|a\right|^{2}-\frac{\left|b\right|^{2}}{2^{n+1}-1}\right)\left|0_{1+n}\right\rangle \left\langle 0_{1+n}\right|,
\end{array}\label{Gl 4.1.1}
\end{equation}
leading us to the Gram-Schmidt-orthonormalized attractor (sub-)space
with respect to the eigenvalue $\lambda=1$ 
\begin{equation}
\left|0_{1+n}\right\rangle \left\langle 0_{1+n}\right|,\,\left(\hat{I}_{1+n}-\left|0_{1+n}\right\rangle \left\langle 0_{1+n}\right|\right)\cdot\left(2^{n+1}-1\right)^{-1/2}.\label{Gl 4.1.2}
\end{equation}

\begin{figure}[H]
\center\includegraphics[scale=0.055]{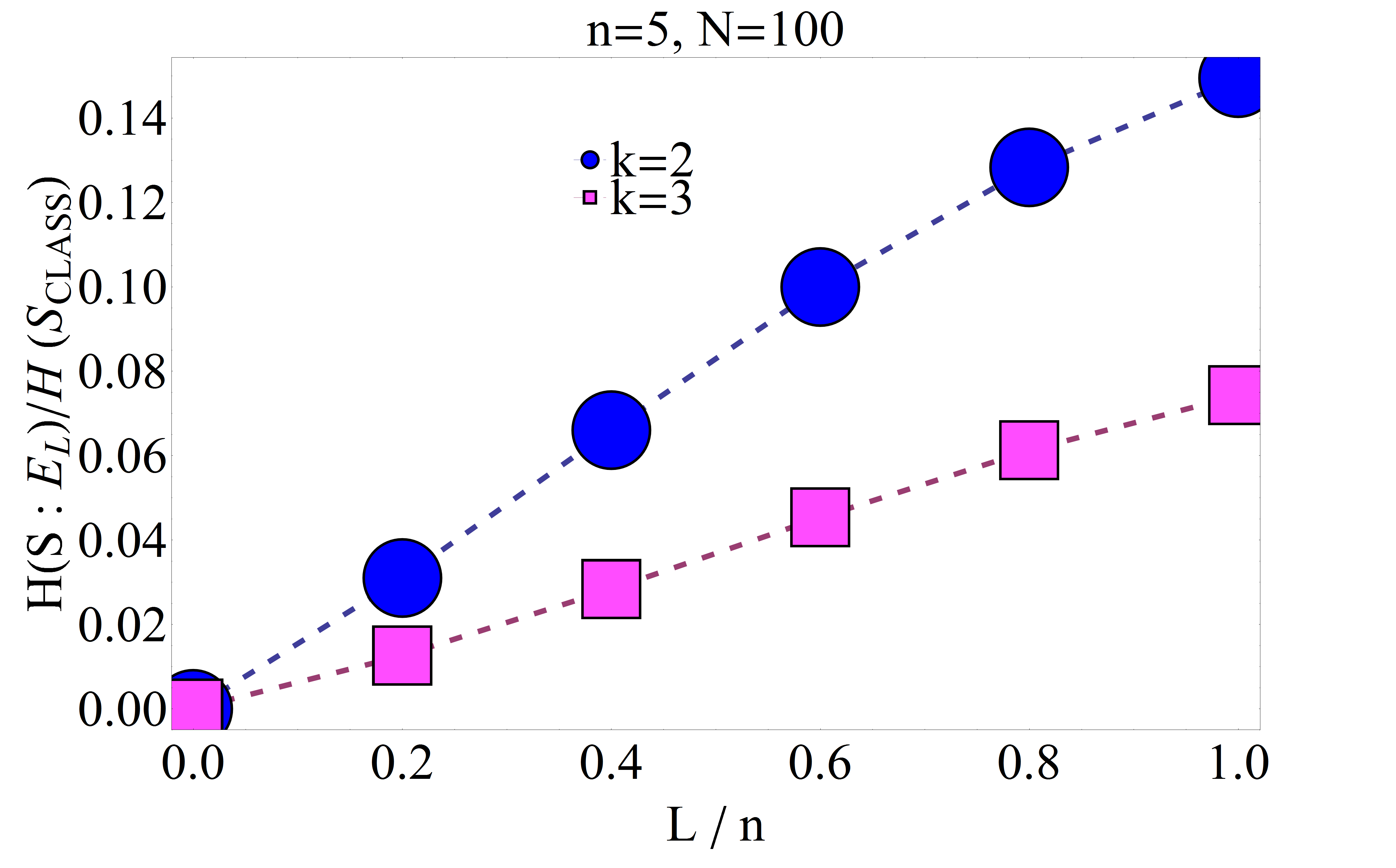}
\caption{Random unitary evolution model: MI $H\left(S:\, E_{L}\right)/H\left(S_{class}\right)$
vs $0<f\leq1$ for $\alpha=\pi/2$ with respect to $\hat{\rho}_{SE}^{\textrm{in}}$
from Fig. \ref{Fig.1} with $N=100$, $k=2,\,3$, $p_{i}=2^{-k}\,\forall\, i\in\left\{ 0,\,...,\,2^{k}-1\right\} $
and $H\left(S_{\textrm{class}}\right)=k$, evolving iteratively by Eq. (\ref{Gl 2.1})-(\ref{Gl 2.5}). \label{Fig.6}}
\end{figure}
Eq. (\ref{Gl 4.1.2}) contains the $\left\{ \lambda=1\right\} -$eigenstate $\left|0_{1+n}\right\rangle $
and $\hat{I}_{1+n}$ as the trivial fixed point state of the random
unitary evolution. Eq. (\ref{Gl 4.1.2}) will appear $\forall\alpha>0$,
even if $\alpha\ll1$, in which case we would need a higher number
$N$ of iterations compared with the $\left(\alpha=\pi/2\right)$-choice.
We can confirm Eq. (\ref{Gl 4.1.2}) by decomposing $\hat{\rho}_{SE}^{\textrm{in}}$
from Fig. \ref{Fig.1} with $\hat{\rho}_{E}^{\textrm{in}}=\left|1_{n}\right\rangle \left\langle 1_{n}\right|$
in accord with Eq. (\ref{Gl 2.7}), that by means of Eq. (\ref{Gl 4.1.2})
yields the asymptotic state
\begin{equation}
\label{Gl 4.1.3}
\begin{array}{l}
\hat{\rho}_{SE}^{\textrm{out}}\left(N\gg1\right)=\underset{\lambda=1,\, i=1}{\overset{2}{\sum}}\lambda^{N}\textrm{Tr}\left\{ \widehat{\rho}_{SE}^{\textrm{in}}\widehat{X}_{\lambda,i}^{\dagger}\right\} \widehat{X}_{\lambda,i}\\
=\frac{\left(\hat{I}_{1+n}-\left|0_{1+n}\right\rangle \left\langle 0_{1+n}\right|\right)}{\left(2^{n+1}-1\right)},
\end{array}
\end{equation}
whose asymptotic PIP coincides exactly with the numerical PIP-plot
in Fig. \ref{Fig.3} (the $\blacksquare$-dotted curve). Eq. (\ref{Gl 4.1.3}) yields again
\[
\underset{N\gg1}{\lim}H\left(S:\, E_{L=n}\right)/H\left(S_{\textrm{class}}\right)=0.
\]

Furthermore, for $L=n\gg1$ Eq. (\ref{Gl 4.1.1}) leads in the standard computational basis (for instance
with $\left|b\right|^{2}=\left|a\right|^{2}=1/2\Rightarrow H\left(S_{\textrm{class}}\right)=1$)
to
\begin{equation*}
\begin{array}{l}
H\left(S\right)\approx 0.811<H\left(S_{\textrm{class}}\right)=1\\
H\left(E_{L=n}\right)\approx H\left(S_{\textrm{class}}\right)+n\left|b\right|^{2}\\
H\left(S,\, E_{L=n}\right)\approx H\left(S_{\textrm{class}}\right)+\left(n+1\right)\left|b\right|^{2}\\
\Rightarrow H\left(S:\, E_{L=n}\right)\approx H\left(S\right)-\left|b\right|^{2}=0.311,
\end{array}
\end{equation*}
as in Fig. \ref{Fig.3} ($\bullet$-dotted curve). Computing $H\left(S:\, E_{L=n}\right)$
for different $n$-values yields results in Table \ref{Tab1}, which
numerically confirms that we need at least $n\geq5$ $E$-qubits to
approximately obtain the correct $H\left(S:\, E_{L=n}\right)$-maximum
in Fig. \ref{Fig.3} ($\bullet$-dotted curve) for $N\gg1$, as in
\cite{key-6}.
\begin{table}[H]
\center%
\begin{tabular}{|c|c|}
\hline 
$H\left(S:\, E_{L=n}\right)/H\left(S_{\textrm{class}}\right)$ & $n$\tabularnewline
\hline 
\hline 
0.124 & 2\tabularnewline
\hline 
0.190 & 3\tabularnewline
\hline 
0.236 & 4\tabularnewline
\hline 
0.266 & 5\tabularnewline
\hline 
0.285 & 6\tabularnewline
\hline 
0.296 & 7\tabularnewline
\hline 
0.303 & 8\tabularnewline
\hline 
0.306 & 9\tabularnewline
\hline 
0.308 & 10\tabularnewline
\hline 
\end{tabular}

\caption{Mutual information (MI) vs the number $n$ of $E$-qubits for
$\hat{\rho}_{SE}^{\textrm{out}}\left(N\gg1\right)$ in Fig. \ref{Fig.3} ($\bullet$-dotted
curve). \label{Tab1}}
\end{table}

Finally, decomposing the initial states $\hat{\rho}_{SE}^{\textrm{in}}$ from the $\blacklozenge$-dotted
and the $\blacksquare$-dotted curve in Fig. \ref{Fig.3} in accord
with Eq. (\ref{Gl 2.7}) we obtain the resulting asymptotic states $\hat{\rho}_{SE}^{\textrm{out}\left(1\right)}$
and $\hat{\rho}_{SE}^{\textrm{out}\left(2\right)}$,
\begin{equation}
\label{Gl 4.1.5}
\begin{array}{l}
\hat{\rho}_{SE}^{\textrm{out}\left(1\right)}\left(N\gg1\right)=\frac{\left|a\right|^{2}}{2}\left|0_{1+n}\right\rangle \left\langle 0_{1+n}\right|+\\
\frac{\left|b\right|^{2}}{2}\left(\hat{I}_{1+n}-\left|0_{1+n}\right\rangle \left\langle 0_{1+n}\right|\right)\cdot\left(2^{n+1}-1\right)^{-1}\\
\hat{\rho}_{SE}^{\textrm{out}\left(2\right)}\left(N\gg1\right)=2^{-n}\left|a\right|^{2}\left|0_{1+n}\right\rangle \left\langle 0_{1+n}\right|+\\
\frac{\left(1-2^{-n}\left|a\right|^{2}\right)}{\left(2^{n+1}-1\right)}\left(\hat{I}_{1+n}-\left|0_{1+n}\right\rangle \left\langle 0_{1+n}\right|\right),
\end{array}
\end{equation}
respectively, whose PIPs coincide for $N\gg1$ exactly with
those in Fig. \ref{Fig.3} ($\blacklozenge$- and $\blacksquare$-dotted
curves).
Thus, the analytic expression $\hat{\rho}_{SE}^{\textrm{out}\left(2\right)}$
for the $\blacksquare$-dotted curve of Fig. \ref{Fig.3} in Eq. (\ref{Gl 4.1.5})
exactly coincides with the analytic expression for the random unitarily
evolved Eq. (\ref{Gl 4.1.1_ent}) with respect to $N\gg1$, as already
confirmed numerically. This validates the anticipated dissipative
attractor (sub-)space in Eq. (\ref{Gl 4.1.2}) with respect to $\lambda=1$
and the $k=1$ qubit system $S$.

It remains to anticipate the $\left\{ \lambda=1\right\} $-attractor
(sub-)space for the $k>1$ qubit system $S$ and its $n\geq k$ qubit
environment $E$: from Fig. \ref{Fig.6} we see that the resulting asymptotic
states $\hat{\rho}_{SE}^{\textrm{out}}\left(N\gg1\right)$ for $k=2,\,3$
may be decomposed as
\begin{equation}
\label{Gl 4.1.6}
\begin{array}{l}
\hat{\rho}_{SE}^{\textrm{out}}\left(N\gg1\right)=\underset{\lambda=1,\, i=1}{\overset{2}{\sum}}\lambda^{N}\textrm{Tr}\left\{ \widehat{\rho}_{SE}^{\textrm{in}}\widehat{X}_{\lambda,i}^{\dagger}\right\} \widehat{X}_{\lambda,i}\\
=\left\langle 0_{k}\right|\hat{\rho}_{S}^{\textrm{in}}\left|0_{k}\right\rangle \left|0_{k+n}\right\rangle \left\langle 0_{k+n}\right|+\\
\frac{\left(1-\left\langle 0_{k}\right|\hat{\rho}_{S}^{\textrm{in}}\left|0_{k}\right\rangle \right)}{\left(2^{n+k}-1\right)} \left(\hat{I}_{k+n}-\left|0_{k+n}\right\rangle \left\langle 0_{k+n}\right|\right),
\end{array}
\end{equation}
with $\left\langle 0_{k}\right|\hat{\rho}_{S}^{\textrm{in}}\left|0_{k}\right\rangle :=p_{0}=\left|a_{0}\right|^{2}=2^{-k}$. Thus, Eq. (\ref{Gl 4.1.6})
forces us to conjecture
\begin{equation}
\left|0_{k+n}\right\rangle \left\langle 0_{k+n}\right|,\,\left(\hat{I}_{k+n}-\left|0_{k+n}\right\rangle \left\langle 0_{k+n}\right|\right)\cdot\left(2^{n+k}-1\right)^{-1/2}\label{Gl 4.1.7}
\end{equation}
as the appropriate $k\geq1$ generalization of the attractor (sub-)space
in Eq. (\ref{Gl 4.1.2}). We can confirm Eq. (\ref{Gl 4.1.7})
by approximating $H\left(S:\, E_{L=n}\right)$ of Eq. (\ref{Gl 4.1.6})
(with $H\left(S_{\textrm{class}}\right)=k$) with respect to $n\sim k\gg1$,
yielding with $\ln\left(1\pm x\right)\approx\pm x+\mathcal{O}\left(x^{2}\right)$
relations
\begin{equation}
\label{Gl 4.1.8}
\begin{array}{l}
H\left(S:\, E_{L=n}\right)\approx k\cdot2^{-k}\\
\Rightarrow H\left(S:\, E_{L=n}\right)/H\left(S_{\textrm{class}}\right)\approx k\cdot2^{-k}/k=2^{-k},
\end{array}
\end{equation}
in accord with Fig. \ref{Fig.6} and \cite{key-6}. Thus, Eq. (\ref{Gl 4.1.8})
validates Eq. (\ref{Gl 4.1.7}), which we want to reproduce analytically in the following
subsection by utilizing the eigenvalue Eq. (\ref{Gl 2.6}).

\textbf{Summary}

Numerical analysis of the random unitary evolution algorithm yields, in presence of symmetric dissipation, a two-dimensional attractor space (dominated by a completely mixed attractor $S$-$E$ state $\hat{I}_{k+n}$) which in general leads to almost completely mixed $\hat{\rho}_{SE}^{\textrm{out}}$. This attractor space structure accounts for the observed loss (leakage) of information about system's pointer basis out of its environment $E$. We will confirm Eq. (\ref{Gl 4.1.7}) in the next subsection.

\subsection{Analytic reconstruction of the symmetric dissipative attractor space\label{A4.2}}

~~~Finally, we want to analytically confirm the numerically reconstructed
dissipative attractor space of the random unitary model. Therefore,
we start with Eq. (\ref{Gl 2.6}) that contains $g:=\left|M\right|\cdot2^{2\left(k+n\right)}$
equations for $2^{2\left(k+n\right)}$ unknown $\widehat{X}_{\lambda,i}$-matrix
entries with respect to the fixed $\lambda$-eigenvalue. We first
reformulate Eq. (\ref{Gl 2.6}) as a linear system of equations
\begin{equation}
A\mathbf{x}=\mathbf{0},\label{Gl 4.2.1}
\end{equation}
with a $\left(g\times g\right)$-matrix $A$ and a $\left(1\times g\right)$-column
vector $\mathbf{x}$ containing the first $2^{2\left(k+n\right)}$
unknown $\widehat{X}_{\lambda,i}$-matrix entries and the remaining
$g^{2}-2^{2\left(k+n\right)}$ zero entries. Then we can apply the
QR-decomposition (s. for instance \cite{key-9}) to $A$ in Eq. (\ref{Gl 4.2.1})
and determine from the rank $r$ of the corresponding $R$-matrix
the dimensionality
\begin{equation}
d_{n\geq k}^{\lambda}=2^{2\left(k+n\right)}-r\label{Gl 4.2.2}
\end{equation}
of the attractor (sub-)space in Eq. (\ref{Gl 2.7}) for a fixed $\lambda$-value,
whereas the $Q$-matrix leads to all allowed $d_{n\geq k}^{\lambda}$
configurations of $\widehat{X}_{\lambda,i}$-matrix entries in Eq. (\ref{Gl 2.7}).
It can be explicitly shown that the QR-decomposition of Eq. (\ref{Gl 4.2.1})
leads for $\lambda=1$ of $\hat{U}_{ij}^{\textrm{Tot}}\left(\alpha\right)$
to $d_{n\geq k}^{\lambda=1}=2$ and diagonal $\widehat{X}_{\lambda=1,i}$-matrix
configurations 
\[\widehat{X}_{\lambda=1,i}=\textrm{diag}[c_{1},\,c_{2},\,...,\,c_{2}]_{2^{k+n}\times2^{k+n}}\]
in the standard computational basis with respect to a $k\geq1$ qubit $S$, with one arbitrary entry $c_{1}$ and $\left(2^{k+n}-1\right)$ identical
entries $c_{2}=const$. Since we know that $\left|0_{k+n}\right\rangle \left\langle 0_{k+n}\right|$
is an eigenstate for $\lambda=1$ of $\hat{U}_{ij}^{\textrm{Tot}}\left(\alpha\right)$,
we first may set, as one of the two
allowed attractor state configurations, $(c_{1}\neq0,\,c_{2}=0)$. The second linearly
independent configuration is certainly the unity matrix $\hat{I}_{k+n}$
as the standard fixed point state of the random unitary dynamics,
given by $\left(c_{1}=c_{2}\neq0\right)$. This confirms
the attractor (sub-)space for $\lambda=1$ of $\hat{U}_{ij}^{\textrm{Tot}}\left(\alpha\right)$
in Eq. (\ref{Gl 4.1.7}) above. On the other hand, for all three remaining
eigenvalues $\lambda\neq1$ of $\hat{U}_{ij}^{\textrm{Tot}}\left(\alpha\right)$
we obtain by means of the QR-decomposition for all $\left(0<\alpha\leq\pi/2\right)$ \[d_{n\geq k}^{\lambda\neq1}=0,\]
in accord with numerical results from subsection \ref{A4.1}.

\textbf{Summary}

Analytic analysis has confirmed, by means of the QR-decomposition, the attractor space structure from Eq. (\ref{Gl 4.1.7}). In the next section \ref{A6} we will therefore turn our attention to the random unitary and Zurek's evolution qubit model in presence of asymmetric dissipation and dephasing and compare the behavior of quantum Darwinism under such modified conditions with results obtained so far.
\section{Random unitary operations perspective on quantum Darwinism: pure
decoherence vs asymmetric dissipation and dephasing \label{A6}}

~~~In this section we investigate the behavior of quantum Darwinism with respect to asymmetric dissipation and dephasing in Zurek's (subsection \ref{A5.1}) and the random unitary model (subsections \ref{A5.2}-\ref{A5.3}). We will notice that in both models asymmetric dissipation suppresses quantum Darwinism, whereas dephasing does not influence the dynamics of an open system and its environment.

\subsection{Pure decoherence, dissipation and dephasing in Zurek's model of quantum
Darwinism\label{A5.1}}
~~~Let us first consider the following cases for Zurek's model
of quantum Darwinism:

\textbf{I) Case $\alpha_{1}\neq\alpha_{2}>0$, $\gamma=0$
(asymmetric dissipation)}: 

This parameter choice generalizes the analysis
of the dissipative qubit model in section \ref{A4}. Setting
without loss of generality $\alpha_{1}=2\pi/3$ and $\alpha_{2}=\pi/3$
in Eq. (\ref{Gl 2.1})-(\ref{Gl 2.4}) we obtain from Zurek's evolution,
for $L=n=2$,
\begin{equation}
\begin{array}{l}
\left|\Psi_{SE}^{\textrm{out}}\right\rangle =\left(A+B+C\right)^{n}\left(\left|\Psi_{S}^{\textrm{in}}\right\rangle \otimes\left|0_{n}\right\rangle \right)= -\frac{a}{2}\left|0\right\rangle \left|01\right\rangle \\
+a\left[\frac{3}{4}\left|0\right\rangle +i\frac{\sqrt{3}}{4}\left|1\right\rangle \right]\left|00\right\rangle +b\left[i\frac{\sqrt{3}}{2}\left|0\right\rangle -\frac{1}{2}\left|1\right\rangle \right]\left|10\right\rangle \\
\left|\Psi_{SE}^{\textrm{out}\left(R\right)}\right\rangle =\hat{U}_{ij}^{\textrm{Diss}}\left(\alpha_{1},\,\alpha_{2},\,\gamma\right)\hat{U}_{ij}^{\left(\phi=\pi/2\right)}\left(\left|\Psi_{S}^{\textrm{in}}\right\rangle \otimes\left|0_{n}\right\rangle \right)\\
=\left(A_{R}+B_{R}+C_{R}+D_{R}\right)^{n}\left(\left|\Psi_{S}^{\textrm{in}}\right\rangle \otimes\left|0_{n}\right\rangle \right)\\
=\left[\frac{3}{4}a+i\frac{\sqrt{3}}{4}b\right]\left|0\right\rangle \left|00\right\rangle +\left[-\frac{1}{4}a+i\frac{\sqrt{3}}{4}b\right]\left|0\right\rangle \left|10\right\rangle \\
+\left[-\frac{1}{4}b+i\frac{\sqrt{3}}{4}a\right]\left|1\right\rangle \left|01\right\rangle +\left[\frac{3}{4}b+i\frac{\sqrt{3}}{4}a\right]\left|1\right\rangle \left|11\right\rangle,
\end{array}\label{Gl 5.1}
\end{equation}
where
\begin{equation*}
\begin{array}{l}
A=\left(\sqrt{3}/2\right)\left|0\right\rangle _{i}\left\langle 0\right|\otimes\left|0\right\rangle _{j}\left\langle 0\right|=A_{R}\\
B=i\left|0\right\rangle _{i}\left\langle 1\right|\otimes\left|1\right\rangle _{j}\left\langle 0\right|,\, B_{R}=0.5i\left|0\right\rangle _{i}\left\langle 1\right|\otimes\left|0\right\rangle _{j}\left\langle 0\right|\\
C=0.5i\left|1\right\rangle _{i}\left\langle 0\right|\otimes\left|1\right\rangle _{j}\left\langle 0\right|=C_{R}\\
D_{R}=\left(\sqrt{3}/2\right)\left|1\right\rangle _{i}\left\langle 1\right|\otimes\left|1\right\rangle _{j}\left\langle 1\right|
\end{array}
\end{equation*}
 for $i\in\left\{ 1\right\} $ and $j\in\left\{ 1,\,...,\, n\right\} $
(s. Appendix \ref{AAG}). Both
$\left|\Psi_{SE_{L=n=2}}^{out}\right\rangle $ and $\left|\Psi_{SE_{L=n=2}}^{\textrm{out}\left(R\right)}\right\rangle $
in Eq. (\ref{Gl 5.1}) lead to $H\left(S\right)=-\underset{i=1}{\overset{2}{\sum}}\lambda_{i}\log_{2}\lambda_{i}<H\left(S_{\textrm{class}}\right)$
with
\begin{equation*}
\begin{array}{l}
\lambda_{1/2}=\frac{1}{2}\pm\frac{1}{8}\sqrt{\underset{\neq16\left(\left|a\right|^{2}-\left|b\right|^{2}\right)^{2}}{\underbrace{\left(3\left|a\right|^{2}+4\left|b\right|^{2}\right)^{2}+4\left|a\right|^{2}}}}\\
\lambda_{1}^{R}=\frac{10}{16}\left|a\right|^{2}+\frac{6}{16}\left|b\right|^{2}+\frac{2i\sqrt{3}}{16}\left(a^{*}b-ab^{*}\right)\\
\lambda_{2}^{R}=\frac{10}{16}\left|b\right|^{2}+\frac{6}{16}\left|a\right|^{2}-\frac{2i\sqrt{3}}{16}\left(a^{*}b-ab^{*}\right)
\end{array}
\end{equation*}
 (for $0<\left|a\right|^{2}<1$), respectively, violating Eq. (\ref{Gl 3.5.1})
\cite{key-6}. Nevertheless, the order of $\hat{U}_{ij}^{\textrm{Diss}}\left(\alpha_{1},\,\alpha_{2}\right)=\hat{U}_{ij}^{\textrm{Diss}}\left(\alpha_{1},\,\alpha_{2},\,\gamma=0\right)$
and $\hat{U}_{ij}^{\left(\phi=\pi/2\right)}$ does matter within Zurek's
qubit model, since the value of $H\left(S\right)$
changes when exchanging the order of the dissipative and the CNOT-part
in Eq. (\ref{Gl 2.4}). However, from the point of view of quantum Darwinism,
it is irrelevant within Zurek's qubit model in which order one applies
$\hat{U}_{ij}^{\textrm{Diss}}\left(\alpha_{1},\,\alpha_{2},\,\gamma=0\right)$
and $\hat{U}_{ij}^{\left(\phi=\pi/2\right)}$ in Eq. (\ref{Gl 2.4}) if
$\alpha_{1}\neq\alpha_{2}>0$ (quantum Darwinism always disappears
in this case).

\textbf{II) Case $\alpha_{1}=\alpha_{2}=0$, $\gamma=\pi$
(pure dephasing):} 

Applying Eq. (\ref{Gl 2.4}) to $\hat{\rho}_{SE}^{\textrm{in}}$
associated with (\ref{Gl 3.5}) in accord with Zurek's qubit model of quantum Darwinism, we obtain
$\forall\,\left(0<L\leq n\right)$
\begin{equation}
\label{Gl 5.2}
\begin{array}{l}
\left|\Psi_{SE_{L=n}}^{\textrm{out}}\right\rangle =\hat{U}_{ij}^{\textrm{Tot}}\left(\gamma=\pi\right)\left(\left|\Psi_{S}^{\textrm{in}}\right\rangle \otimes\left|0_{n}\right\rangle \right)=\\
=\left(-i\right)^{n}\left(a\left|0\right\rangle \left|0_{n}\right\rangle +\left(-1\right)^{n}b\left|1\right\rangle \left|1_{n}\right\rangle \right)
\\
\left|\Psi_{SE_{L=n}}^{\textrm{out}\left(R\right)}\right\rangle =\hat{U}_{ij}^{\textrm{Diss}}\left(\gamma=\pi\right)\hat{U}_{ij}^{\left(\phi=\pi/2\right)}
\left(\left|\Psi_{S}^{\textrm{in}}\right\rangle \otimes\left|0_{n}\right\rangle \right)\\
=\left(-i\right)^{n}\left(a\left|0\right\rangle \left|0_{n}\right\rangle +b\left|1\right\rangle \left|1_{n}\right\rangle \right),
\end{array}
\end{equation}
showing that the order of the dephasing and the CNOT-operator in Eq. (\ref{Gl 2.4})
does not matter within Zurek's model of quantum Darwinism: both $\left|\Psi_{SE_{L}}^{\textrm{out}}\right\rangle $
and $\left|\Psi_{SE_{L}}^{\textrm{out}\left(R\right)}\right\rangle $ in Eq. (\ref{Gl 5.2})
lead to quantum Darwinism with a PIP and $H\left(S:\, E_{L}\right)/H\left(S_{\textrm{class}}\right)$
as in Eq. (\ref{Gl 3.5}) and Fig. \ref{Fig.1} ($\bullet$-dotted curve).

\textbf{III) Case $\alpha_{1}=\alpha_{2}=\pi/2$,
$\gamma=\pi$ (dissipation and dephasing):}

For this parameter
choice Eq. (\ref{Gl 2.4}), applied to $\hat{\rho}_{SE}^{\textrm{in}}$ associated with Eq. (\ref{Gl 3.5})
in accord with Zurek's qubit model of quantum Darwinism, yields 
\begin{equation}
\left|\Psi_{SE_{L=n}}^{\textrm{out}}\right\rangle =\left(-i\right)^{n}\left(a\left|0\right\rangle \left|0_{n}\right\rangle -ib\left|0\right\rangle \left|10_{n-1}\right\rangle \right),\label{Gl 5.3}
\end{equation}
 i.e. quantum Darwinism disappears, regardless of how one exchanges the dissipative
and dephasing operators in $\hat{U}_{ij}^{\textrm{Tot}}$. On the other hand, exchanging
the order of the dissipative-dephasing part and $\hat{U}_{ij}^{\left(\phi=\pi/2\right)}$
in $\hat{U}_{ij}^{\textrm{Tot}}$ yields within Zurek's model of quantum Darwinism 
\begin{equation}
\label{Gl 5.4}
\begin{array}{l}
\left|\Psi_{SE_{L=n}}^{\textrm{out}\left(R\right)}\right\rangle =\hat{U}_{ij}^{\textrm{Diss}}\left(\alpha_{1},\,\alpha_{2},\,\gamma\right)\hat{U}_{ij}^{\left(\phi=\pi/2\right)}\left(\left|\Psi_{S}^{\textrm{in}}\right\rangle \otimes\left|0_{n}\right\rangle \right)\\
=\left(-i\right)^{n}\left(a\left|0\right\rangle \left|0_{n}\right\rangle +b\left|1\right\rangle \left|1_{n}\right\rangle \right),
\end{array}
\end{equation}
 i.e. quantum Darwinism appears, regardless of how one exchanges the dissipative
and dephasing operators among each other. This is expected, since
$\alpha_{1}=\alpha_{2}=\pi/2$ corresponds to the symmetric dissipative
operator studied in the previous section \ref{A4}.
\\
\textbf{Summary}

There are to important conclusions which can be drawn from the above discussion of Zurek's qubit evolution model:
\begin{itemize}
\item Asymmetric dissipation always suppresses quantum Darwinism, regardless of the order in which the CNOT and the asymmetric operation from Eq. (\ref{Gl 2.4}) are applied to a given inital $S$-$E$ state.
\item The dephasing part of Eq. (\ref{Gl 2.4}) does not influence the dynamics of initial states at all.
\end{itemize}
In the next subsection we will perform a numerical analysis of quantum Darwinism subjected to asymmetric dissipation with dephasing in the framework of the random unitary evolution model.

\subsection{Random unitary model: Numerical reconstruction of the dissipative-dephased attractor space\label{A5.2}}

~~~As in the last subsection, we again concentrate on the following three cases, this time with respect to the random unitary model: 

\textbf{I) Case $\alpha_{1}\neq\alpha_{2}>0$, $\gamma=0$
(asymmetric dissipation)}: 

We set, without loss of generality, \[\alpha_{1}=2\pi/3\neq\alpha_{2}=\pi/3;\]
for this parameter choice Eq. (\ref{Gl 2.4}) has eigenvalues \[\lambda_{1/2}=\pm1\\,\,\,\lambda_{3/4}=0.43\pm i\cdot0.9\]
and we see that Eq. (\ref{Gl 2.4}),
applied to $\hat{\rho}_{SE}^{\textrm{in}}$ in Fig. \ref{Fig.1} in accord
with Eq. (\ref{Gl 2.1})-(\ref{Gl 2.5}), leads $\forall\left(\alpha_{1}\neq\alpha_{2}>0\right)$
to the PIP given by the $\blacksquare$-dotted curve in Fig. \ref{Fig.3}
above (with $n=8$, $k=1$ and $N\gg1$). 

The same PIP emerges $\forall\left(\alpha_{1}\neq\alpha_{2}>0\right)$
in the limit $N\gg1$ if one starts the random unitary evolution with
$\hat{\rho}_{SE}^{\textrm{in}}$ from Fig. \ref{Fig.1}, whose $\hat{\rho}_{S}^{\textrm{in}}$
represents a pure, $k>1$ qubit input-system $S$: we again see that
$H\left(S:\, E_{L}\right)=0$ $\forall\left(0\leq L\leq n\right)$.
When looking at the corresponding final (stationary) state 
\begin{equation}
\hat{\rho}_{SE}^{\textrm{out}}=2^{-\left(k+n\right)}\cdot\hat{I}_{k+n}\label{Gl 5.5}
\end{equation}
from Fig. \ref{Fig.3} ($\blacksquare$-dotted curve) for $n\gg k\geq1$,
we are tempted to conclude, by means of Eq. (\ref{Gl 2.7}), that the
attractor space capable of reproducing Eq. (\ref{Gl 5.5}) should be associated
with the eigenvalue $\left\{ \lambda=1\right\} $ of the unitary transformation $\hat{U}_{e}^{\textrm{Tot}}\left(\alpha_{1}=2\pi/3=2\alpha_{2},\,\gamma=0\right)$
from Eq. (\ref{Gl 2.4}) and display the structure
\begin{equation}
\hat{X}_{\lambda=1,\, i=1}=2^{-\left(k+n\right)/2}\cdot\hat{I}_{k+n},\label{Gl 5.6}
\end{equation}
i.e. $d_{n\gg k}^{\lambda=1}=1$ (dimensionality one). This would
imply that asymmetric dissipation yields completely mixed final (stationary) $S$-$E$-states from Eq. (\ref{Gl 5.6}) with 
\begin{equation}
H\left(S,\, E_{L}\right)=H\left(S\right)+H\left(E_{L}\right)=H\left(S_{class}\right)+
L=k+L\label{Gl 5.7}
\end{equation}
$\forall\left(0\leq L\leq n\right)$, implying $\forall\left(0\leq L\leq n\right)$ indeed the vanishing MI, $H\left(S:\, E_{L}\right)=0$, as suggested by the $\blacksquare$-dotted
curve in Fig. \ref{Fig.3}. We will confirm Eq. (\ref{Gl 5.6}) analytically
in subsection \ref{A5.3}, however, before doing that we want to mention
another interesting issue: starting the random unitary evolution of
$\hat{\rho}_{SE}^{\textrm{in}}$ from Fig. \ref{Fig.1} with one of the two
dissipative parameters set to zero, for instance with $\alpha_{1}=2\pi/3\neq\alpha_{2}=0=\gamma$,
we would again obtain in the limit $N\gg1$ a PIP given by the $\blacksquare$-dotted
curve in Fig. \ref{Fig.3} above.

Thus, in the asymptotic limit $N\gg1$, and this is the most significant
conclusion, both PIPs, for $\alpha_{1}=2\pi/3\neq\alpha_{2}=\pi/3\neq\gamma=0$
as well as for $\alpha_{1}=2\pi/3\neq\alpha_{2}=0=\gamma$, tend to
zero $\forall\,0\leq L\leq n$ according to the $\blacksquare$-dotted
curve in Fig. \ref{Fig.3}. This means that for $N\gg1$ already the
fact that at least one of the dissipative parameters does not vanish
suffices to obtain completely mixed final (stationary) $S$-$E$-states such as
the one in Eq. (\ref{Gl 5.5}) and the corresponding $\left\{ \lambda=1\right\} $
attractor subspace from Eq. (\ref{Gl 5.6}). We assume Eq. (\ref{Gl 5.6}) to be the only
eigenvalue subspace $\mathcal{A}_{\lambda=1}$ of non-zero dimension
contributing to the entire attractor space $\mathcal{A}$ in (\ref{Gl 2.7})
above.

On the other hand, we also may point out that the $\blacksquare$-dotted
curve in Fig. \ref{Fig.3} indicates the negative impact which asymmetric
dissipation has on quantum Darwinism. Namely, asymmetric values of $\alpha_{1}$ and
$\alpha_{2}$ diminish in general the dimensionality of the $2$-dimensional
$\left\{ \lambda=1\right\} $ attractor subspace emerging from a symmetric
($\alpha_{1}=\alpha_{2}=\alpha$) dissipative random unitary evolution
discussed in section \ref{A4} and in subsection \ref{A5.3}
below. This enables us to store a higher amount of $H\left(S_{\textrm{class}}\right)$
in the limit $N\gg1$ (namely $0.3\cdot H\left(S_{class}\right)$
according to the $\bullet$-dotted curve in Fig. \ref{Fig.3})
if we symmetrize the dissipative contributions in $\hat{U}_{e}^{\textrm{Tot}}$
of Eq. (\ref{Gl 2.4}), instead of setting $\alpha_{1}\neq\alpha_{2}$, which leads to $\underset{N\gg1}{\lim}H\left(S:\, E_{L=n}\right)=0$,
as indicated by the $\blacksquare$-dotted curve in Fig. \ref{Fig.3}).

All above conclusions for this parameter choice in the limit $N\gg1$
remain unchanged if we exchange the order of application of $\hat{U}_{ij}^{\left(\phi=\pi/2\right)}$
and $\hat{U}_{ij}^{\textrm{Diss}}\left(\alpha_{1}=2\pi/3=2\alpha_{2},\,\gamma=0\right)$
in Eq. (\ref{Gl 2.4}), as was the case for the completely symmetric ($\alpha_{1}=\alpha_{2}=\alpha$)
dissipative random unitary evolution discussed in section \ref{A4}. 

\textbf{II) Case $\alpha_{1}=\alpha_{2}=0$, $\gamma\neq0$
(pure dephasing):}

We set, without loss of generality, $\alpha_{1}=\alpha_{2}=0\neq\gamma=\pi$:
for this parameter choice Eq. (\ref{Gl 2.4}) has eigenvalues \[\lambda_{1/2}=\pm1,\,\,\lambda_{3/4}=\pm i\] and we see that Eq. (\ref{Gl 2.4}), applied
to $\hat{\rho}_{SE}^{\textrm{in}}$ in Fig. \ref{Fig.1} in accord with Eq. (\ref{Gl 2.1})-(\ref{Gl 2.5}),
leads $\forall\gamma>0$ in the limit $N\gg1$ to the PIP given by
the $\blacksquare$-dotted curve in Fig. \ref{Fig.1} (with $n=8$
and $k=1$). In other words, the attractor space structure of the
maximal attractor space associated with pure decioherence (s. \cite{key-6})
remains unchanged if we add dephasing to the random unitary evolution
with pure decoherence (given by the parameter fixture $\alpha_{1}=\alpha_{2}=0=\gamma$).

Indeed, when looking at the eigenspectrum of $\hat{U}_{e}^{\textrm{Tot}}$
for this specific parameter choice, we may anticipate that eigenvalues
$\lambda_{1/2}=\pm1$ are precisely associated with the corresponding
attractor subspaces emerging from the random unitary evolution with
pure decoherence. On the other hand, the attractor subspaces associated with
eigenvalues $\lambda_{3/4}=\pm i$ of $\hat{U}_{e}^{\textrm{Tot}}\left(\alpha_{1}=\alpha_{2}=0\neq\gamma=\pi\right)$
may be assumed to be zero-dimensional $\forall\gamma>0$, i.e. they do
not contribute to the entire attractor space $\mathcal{A}$ from Eq. (\ref{Gl 2.7})
above. This is reasonable, since $\hat{U}_{e}^{\textrm{Diss}}\left(\gamma\right)$
in $\hat{U}_{e}^{\textrm{Tot}}$ of Eq. (\ref{Gl 2.4}) would only iteratively
change the phase in certain addends of the corresponding resulting final (stationary) state
$\hat{\rho}_{SE}^{\textrm{out}}$, which should, however, not influence the
asymptotic $N\gg1$ evolution behavior of $\hat{\rho}_{SE}^{\textrm{in}}$
and its attractor space. We will confirm these assumptions analytically
in the forthcoming subsection \ref{A5.3}.

Also, all above conclusions for this parameter choice in the limit
$N\gg1$ remain unchanged if we exchange the order of application of $\hat{U}_{ij}^{\left(\phi=\pi/2\right)}$
and $\hat{U}_{ij}^{\textrm{Diss}}\left(\alpha_{1}=\alpha_{2}=0,\,\gamma\neq0\right)$
in Eq. (\ref{Gl 2.4}).

\textbf{III) Case $\alpha_{1}=\alpha_{2}=\pi/2$, $\gamma=\pi$
(dissipation and dephasing):} 

Taking the above results into
account we may expect that the presence of dephasing should not affect
the results of the random unitary evolution with respect to the symmetric dissipation
obtained in section \ref{A4}. Indeed, if we set, without
loss of generality, $\alpha_{1}=\alpha_{2}=\pi/2=\gamma/2$, then
for this parameter choice Eq. (\ref{Gl 2.4}) has eigenvalues \[\lambda_{1/2}=\pm i,\,\,\lambda_{3/4}=-\exp\left[\pm i\cdot\pi/6\right].\] Furthermore, we see
that Eq. (\ref{Gl 2.4}), applied to $\hat{\rho}_{SE}^{\textrm{in}}$ in Fig. \ref{Fig.1}
in accord with Eq.(\ref{Gl 2.1})-(\ref{Gl 2.5}), leads in the limit
$N\gg1$ $\forall\left(\alpha,\,\gamma>0\right)$ to the PIP given
by the $\bullet$-dotted curve in Fig. \ref{Fig.3} above (with
$n=8$ and $k=1$).

The PIP given by the $\bullet$-dotted curve in Fig. \ref{Fig.3}
coincides in the limit $N\gg1$ exactly with the PIP obtained in section \ref{A4} with respect to the symmetric dissipative attractor space without
dephasing. Apparently, the dephasing part in $\hat{U}_{e}^{Tot}\left(\alpha_{1}=\alpha_{2}=\pi/2=\gamma/2\right)$
of Eq. (\ref{Gl 2.4}) does not change the asymptotic (long time) behavior
of the MI with respect to $f=L/n$. By looking at the corresponding
resulting final (stationary) state $\hat{\rho}_{SE}^{\textrm{out}}$ emerging from the PIP given
by the $\bullet$-dotted curve in Fig. \ref{Fig.3} in the limit
$N\gg1$ of the random unitary evolution, we see that it is exactly
the same as $\hat{\rho}_{SE}^{\textrm{out}}$ obtained from the asymptotic
random unitary evolution with $\hat{U}_{e}^{\textrm{Tot}}\left(\alpha_{1}=\alpha_{2}=\pi/2\neq\gamma=0\right)$
of Eq. (\ref{Gl 2.4}). Also, all
above conclusions for this parameter choice in the limit $N\gg1$
remain unchanged if we exchange the order of application of $\hat{U}_{ij}^{\left(\phi=\pi/2\right)}$
and $\hat{U}_{ij}^{\textrm{Diss}}\left(\alpha_{1}=\alpha_{2}\neq0,\,\gamma\neq0\right)$
in Eq. (\ref{Gl 2.4}) $\forall\alpha,\,\gamma>0$.

This leads us to the conclusion that dephasing does not influence
the (dis-)appearance of quantum Darwinism in the random unitary model. On the contrary, quantum
Darwinism appears to depend in the course of the unitary transformation $\hat{U}_{e}^{\textrm{Tot}}\left(\alpha_{1}=\alpha_{2}=\pi/2=\gamma/2\right)$
of Eq. (\ref{Gl 2.4}) only on the interplay between the dissipative and
the pure decoherence part: in other words, in the random unitary model
quantum Darwinism appears only if in $\hat{U}_{e}^{\textrm{Tot}}$ of Eq. (\ref{Gl 2.4})
we have \textbf{$\alpha_{1}=\alpha_{2}=0\leq\gamma$}, otherwise the
MI-'plateau' remains suppressed as soon as $\alpha_{1}\neq0$, $\alpha_{2}\neq0$
or $\left(\alpha_{1},\,\alpha_{2}\right)\neq0$, regardless of the
order, in which the CNOT, dissipative and the dephasing part in $\hat{U}_{e}^{\textrm{Tot}}$
of Eq. (\ref{Gl 2.4}) are applied to a given initial state $\hat{\rho}_{SE}^{\textrm{in}}$.
In the framework of Zurek's qubit model we, however, have to be cautious:
it is certainly true that $\hat{U}_{e}^{\textrm{Tot}}\left(\alpha_{1}=\alpha_{2}=0\leq\gamma\right)$
would always lead to quantum Darwinism, regardless of the order, in
which the CNOT and the dephasing part in $\hat{U}_{e}^{\textrm{Tot}}$ of Eq. (\ref{Gl 2.4})
are applied to a given initial state $\hat{\rho}_{SE}^{in}$; on the
other hand, in Zurek's qubit model the order in which the CNOT and
the dissipative part in $\hat{U}_{e}^{\textrm{Tot}}$ of Eq. (\ref{Gl 2.4}) are
applied to a given initial state $\hat{\rho}_{SE}^{\textrm{in}}$ does matter
for $\alpha_{1}=\alpha_{2}=\alpha$. This is why $\hat{U}_{e}^{\textrm{Tot}}\left(\alpha_{1},\,\alpha_{2},\,\gamma\right)$
and the reversed unitary transformation $\hat{U}_{e}^{\textrm{Tot}\left(R\right)}\left(\alpha_{1},\,\alpha_{2},\,\gamma\right)=\hat{U}_{ij}^{\textrm{Diss}}\left(\alpha_{1},\,\alpha_{2},\,\gamma\right)\hat{U}_{ij}^{\left(\phi=\pi/2\right)}$
for (at least one) non vanishing (asymmetric) dissipative parameter
(i.e. $\alpha_{1}\neq0$ or $\alpha_{2}\neq0$ or $\left(\alpha_{1}\neq\alpha_{2}\right)\neq0$)
would not lead to quantum Darwinism, whereas $\hat{U}_{e}^{\textrm{Tot}\left(R\right)}\left(\alpha_{1}=\alpha_{2}=\alpha,\,\gamma\right)$
would allow us to see the MI-'plateau'.

We will confirm the above conclusions regarding the random unitary
iterative dynamics analytically in the forthcoming subsection \ref{A5.3}.

\textbf{Summary}

The above numerical analysis has demonstrated that, as in the course of Zurek's qubit model, in the presence of asymmetric dissipation with dephasing the random unitary evolution remains influenced only by the dissipative part in Eq. (\ref{Gl 2.4}). Regardless of the order in which asymmetric dissipation and the CNOT operation act upon a given initial $S$-$E$ state, quantum Darwinism does not appear in the asymptotic limit of the random unitary evolution. In the next subsection we will confirm these results analytically by explicitly determining the structure of the corresponding attractor spaces.

\subsection{Random unitary model: Analytic reconstruction of the dissipative-dephased attractor space\label{A5.3}}

~~~In the following, we apply the QR-decomposition \cite{key-9}
to the following parameter values in Eq. (\ref{Gl 2.4}) and obtain the
corresponding attractor-spaces:

\textbf{I) Case $\alpha_{1}\neq\alpha_{2}>0$, $\gamma=0$
(asymmetric dissipation)}: 

After solving the eigenvalue Eq. (\ref{Gl 2.6})
for $\lambda=1$ and $n\ge k\ge1$ by means of the QR-decomposition
method, we obtain $\forall\left(\alpha_{1}\neq\alpha_{2}\right)>0$
with respect to the standard computational basis the diagonal attractor state structure
\[\widehat{X}_{\lambda=1,i}=c_{3}\cdot\widehat{I}_{k+n}\]
(with $2^{k+n}$ identical diagonal entries $c_{3}=const$). For eigenvalues $\lambda\neq1$
of $\hat{U}_{e}^{\textrm{Tot}}$ from Eq. (\ref{Gl 2.4}) the eigenvalue Eq.
(\ref{Gl 2.6}) and the method of QR-decomposition yield, however, vanishing attractor
subspaces, i.e. $d_{n\geq k}^{\lambda\neq1}=0$. This diagonal attractor state structure shows that the corresponding attractor subspace is one dimensional,
$d_{n\geq k}^{\lambda=1}=1$, with the single attractor state being
\begin{equation}
\hat{X}_{\lambda=1,\, i=1}=2^{-\left(k+n\right)/2}\cdot\hat{I}_{k+n}.\label{Gl 5.9}
\end{equation}
Indeed, by means of Eq. (\ref{Gl 5.9}) and Eq. (\ref{Gl 2.7}) we are able
to exactly reproduce the output state from Eq. (\ref{Gl 5.5}), which confirms
our assumption from Eq. (\ref{Gl 5.6}) about the $\left\{ \lambda=1\right\} $-attractor
space structure in subsection \ref{A5.2}, derived w.r.t. the parameter
choice $\alpha_{1}=2\pi/3=2\alpha_{2}\neq\gamma=0$.

\textbf{II) Case $\alpha_{1}=\alpha_{2}=0$, $\gamma\neq0$
(pure dephasing): }

Without any loss of generality we again concentrate
on $\hat{U}_{e}^{\textrm{Tot}}$ with $\alpha_{1}=\alpha_{2}=0\neq\gamma=\pi$
and its eigenvalues $\lambda_{1/2}=\pm1$ and $\lambda_{3/4}=\pm i$.
After solving the eigenvalue Eq. (\ref{Gl 2.6}) for $\lambda=\pm1$
and $n\ge k\ge1$ by means of the QR-decomposition method, we obtain
$\forall\gamma\geq0$ the dimensionally maximal $\left\{ \lambda=1\right\} $-
and $\left\{ \lambda=-1\right\} $-attractor subspaces of pure decoherence
(with $\alpha_{1}=\alpha_{2}=0=\gamma$, s. \cite{key-6}), whereas
the QR-decomposition enables us to conclude that Eq. (\ref{Gl 2.6}) for
$\lambda=\pm i$ and $n\ge k\ge1$ yields zero-dimensional attractor
subspaces, $d_{n\ge k}^{\lambda=\pm i}=0$. This is clear if
we recall that the eigenvalue Eq. (\ref{Gl 2.6}) establishes
relations (constraints) between (in general complex-valued) entries
of the pre-configuration matrix of the type
\begin{equation}
a+ib=c+id,\label{Gl 5.10}
\end{equation}
with $a,\, b,\, c,\, d\in\mathbb{R}$. For instance, for $\lambda=\pm1$
Eq. (\ref{Gl 2.6}) would require that $a=c$ and $b=d$ should hold.
W.r.t. $\lambda=\pm i$ Eq. (\ref{Gl 5.10}) would acquire an additional
phase factor when compared with the case $\lambda=\pm1$, leading
to the general constraint
\begin{equation}
a+ib=\left(\pm i\right)\left(a+ib\right),\label{Gl 5.11}
\end{equation}
as a modifed version of Eq. (\ref{Gl 5.10}) weighted by an additional
phase factor $i$. Unfortunately, Eq. (\ref{Gl 5.11}) has only one solution,
namely the trivial one: $a=b=0$ $\forall\gamma\geq0$
and for all entries of the attractor state matrix. This leads to
the trivial solution
\begin{equation}
\hat{X}_{\lambda=\pm i}=\hat{\mathbf{0}}_{2^{k+n}\times2^{k+n}}\label{Gl 5.12}
\end{equation}
of Eq. (\ref{Gl 2.6}) w.r.t. $\lambda=\pm i$ and $\forall\gamma\geq0$,
indicating that the dimensionality of $\left\{ \lambda=\pm i\right\} $-attractor
subspaces is indeed zero. In other words, the dissipative part of
$\hat{U}_{e}^{\textrm{Tot}}$ from Eq. (\ref{Gl 2.4}) does not contribute to the
entire attractor space \[\mathcal{A}:=\underset{\lambda}{\oplus}\mathcal{A}_{\lambda}.\] This confirms conclusions of
subsection \ref{A5.2}.

\textbf{III) Case $\alpha_{1}=\alpha_{2}=\pi/2$, $\gamma=\pi$
(maximal dissipation and dephasing):} 

For $\alpha_{1}=\alpha_{2}=\pi/2=\gamma/2$
Eq. (\ref{Gl 2.4}) has eigenvalues $\lambda_{1/2}=\pm i$ and $\lambda_{3/4}=-\exp\left[\pm i\cdot\pi/6\right]$
and we see that with respect to $\lambda_{1/2}=\pm i$ the eigenvalue Eq.
(\ref{Gl 2.6}) yields attractor subspaces of zero-dimension (as expected,
since the dissipative part of $\hat{U}_{e}^{\textrm{Tot}}$ from Eq. (\ref{Gl 2.4})
does not contribute to the entire attractor space $\mathcal{A}$).
On the other hand, the QR-decomposition reveals that the only attractor subspace of non-zero dimension for the above parameter choice with symmetric dissipation ($\alpha_{1}=\alpha_{2}=\alpha>0$) is still the $\left\{ \lambda=1\right\} $-attractor
subspace from Eq. (\ref{Gl 4.1.7}), despite of $\gamma>0$.

All these facts indicate that even with respect to eigenvalues $\lambda_{3/4}=-\exp\left[\pm i\cdot\pi/6\right]$
of the unitary transformation given by $\hat{U}_{e}^{\textrm{Tot}}\left(\alpha_{1}=\alpha_{2}=\pi/2=\gamma/2\right)$
from Eq. (\ref{Gl 2.4}) only contributions from the $\left\{ \lambda=1\right\} $-attractor
subspace in Eq. (\ref{Gl 4.1.7}) would persist in Eq. (\ref{Gl 2.5}) in the
limit $N\gg1$. This would in turn explain the fact that in the limit
$N\rightarrow\infty$ the $\bullet$-dotted curve of the PIP in Fig. \ref{Fig.3},
emerging from the random unitary evolution with unitary transformation $\hat{U}_{e}^{\textrm{Tot}}\left(\alpha_{1}=\alpha_{2}=\pi/2=\gamma/2\right)$,
exactly coincides with the PIP that would be obtained by means of
a random unitary evolution with unitary transformation given by $\hat{U}_{e}^{\textrm{Tot}}\left(\alpha_{1}=\alpha_{2}=\pi/2\neq\gamma=0\right)$. 

Thus, we may conclude that within the
random unitary model dephasing does not influence the appearance and disappearance
of quantum Darwinism. In other words, we may say that in the limit $N\rightarrow\infty$ of the random
unitary model dephasing is, if present, being >>averaged out<< in
the course of subsequent iterations.
On the other hand, symmetric and asymmetric dissipation suppress the MI-'plateau' in the asymptotic evolution
limit $N\gg1$, regardless of the order in which pure decoherence
(CNOT), dissipation and dephasing within $\hat{U}_{e}^{\textrm{Tot}}$ of Eq. (\ref{Gl 2.4})
are applied to a given initial state $\hat{\rho}_{SE}^{\textrm{in}}$ in accord
with the iteration procedure of Eq. (\ref{Gl 2.5}). 

\textbf{Summary}

The analytic reconstruction of the corresponding attractor spaces of the random unitary evolution in the presence of asymmetric dissipation with dephasing has confirmed the numerical findings of the previous subsection \ref{A5.2}: the asymmetric dissipative attractor space is one-dimensional, containing only a completely mixed attractor state $\hat{I}_{k+n}$. Such attractor state structure inevitably leads to completely mixed final $S$-$E$ states, causing a breakdown of the quantum Darwinistic MI-plateau' in the asymptotic limit of the random unitary evolution.

\textbf{IV) A broader perspective on quantum Darwinism} 

By means of the concept of quantum Darwinism one tries to explain under which conditions interactions between an open system $S$ and its environment $E$ could allow the pointer basis of the former to be reconstructed by the latter. This is indeed not an easy task. After all, it has already been demonstrated that quantum Darwinism is a model dependent physical phenomenon (see \cite{key-6}).

Nevertheless, if one assumes that the system's pointer basis exists, both Zurek's and the random unitary evolution model single out one preferrable structure of the environment: a pure, one-registry (tensor product) input state containing $n$ mutually non-interacting qudits ($2^{k}$-level systems) \cite{key-3,key-6}. This particular environmental input state structure lets "quasi-classical" system's pointer states emerge from the perspective of the environment after the system-environment interactions have entangled $S$ with $E$ in a way that induces effective decoherence and correlates each $S$-pointer state with one unique $n$-qudit $E$-registry state orthogonal with respect to other environmental one-registry "pointer states" ("entanglement monogamy" \cite{key-3,key-0_4}).

However, as soon as dissipation enters into the evolution of $S$ and $E$, the output state of the total system tends to acquire a bi-partite  tensor product structure composed of completely mixed subsystems which suppresses the appearance of quantum Darwinism. This behavior indicates that the concept of quantum Darwinism, despite its merits, needs to be refined in order to account for the appearance of system's pointer states within uncontrollable environments affected by dissipation. Some suggestions regarding further research within the framework of quantum Darwinism and random unitary operations will be given in the following section \ref{A7}.

The explanation of the appearance of "quasi-classical" system's pointer states in the environment is, however, not the only motivation for the quantum Darwinistic approach to the dynamics of open quantum systems: beside the rather academic problem associated with the appearance of Classicality resulting from the dynamics of open quantum systems discussed so far, there is also an intricate and more practical issue of using the environment as an efficient quantum memory for storing information about the system's pointer basis. 

Taking the results of \cite{key-6} as well as the above discussion into account, we may conclude that the one-registry pure input state structure of the $n$ qudit interaction-free environment represents, from the point of view of Zurek's and the iterative random unitary model, the ideal quantum memory for storing  information about the system's pointer basis with highest efficiency. This quantum memory could be realized experimentally for instance by means of optical lattices with one neutral atom trapped at each lattice site (highly controlled environments), as long as one manages to minimize the dissipative loss of information about system's pointer basis and its "leakage" into an uncontrolled environment. 

Finally, some comments on the 2015 Nature Comm. paper \cite{key-0_4} of Brand$\tilde{\textrm{a}}$o et al. are necessary: our results regarding quantum Darwinism do not contradict the fact that, according to \cite{key-0_4}, quantum Darwinism is a generic physical phenomenon. On the contrary, this result of \cite{key-0_4} remains correct, since Brand$\tilde{\textrm{a}}$o et al. mean by the term "generic" that "the central features of quantum Darwinism are [...] consequences only of the basic structure of quantum mechanics" (see the last sentence of the introduction section I in \cite{key-0_4}), without making any specific assumptions regarding the dynamics. In addition, the results obtained in this paper (and in \cite{key-6}) indicate that quantum Darwinism, although based on fundamental mathematical structures and physical assumptions of quantum mechanics, also depends, from the point of view of objectivity of outcomes, on a specific dynamical model being used. 

Indeed, as Brand$\tilde{\textrm{a}}$o et al. point out in their concluding part (section IV) of \cite{key-0_4} when discussing the two main properties of quantum Darwinism - objectivity of observables and objectivity of outcomes: "[...] the first property - the objectivity of observables - is completely general, being a consequence of quantum formalism only (in particular properties related to the monogamy of entanglement, [...]). On the other hand, the validity of objectivity of outcomes does seem to depend on the details of the evolution, [...]." In other words, the structure of the dynamical model also influences the (dis-)appearance of quantum Darwinism and of the objectivity of observable outcomes, in agreement with \cite{key-0_4} and results obtained in this paper. This means that, despite being generic in the sense of \cite{key-0_4}, quantum Darwinism is nevertheless a model-dependent physical phenomenon.

\section{Conclusions and outlook \label{A7}}
~~~In this paper we discussed the influence of dissipation and dephasing on quantum Darwinism in the framework of Zurek's and the random unitary qubit model. We have seen that in both models dissipation in general suppresses the appearance of quantum Darwinism, whereas dephasing does not influence the system-environment dynamics. 

The order of application of CNOT and the dissipative-dephasing operation from Eq. (\ref{Gl 2.3})
in Eq. (\ref{Gl 2.4}) is relevant for the appearance of quantum Darwinism in Zurek's
qubit model in case of symmetric dissipation. Otherwise, in case of asymmetric dissipation
quantum Darwinism does not appear in Zurek's qubit model, regardless of the order
in which one applies CNOT and
Eq. (\ref{Gl 2.3}) within
Eq. (\ref{Gl 2.4}). Only in case of vanishing dissipation
quantum Darwinism appears within Zurek's qubit model, regardless of how one interchanges
CNOT-operations and dephasing-operators
in Eq. (\ref{Gl 2.4}). Thus, dephasing does not influence the system-environment dynamics in Zurek's qubit model of quantum Darwinism.

In case of asymmetric dissipation within the random unitary model the corresponding
attractor space is one dimensional,
with the only attractor state given by Eq. (\ref{Gl 5.9})
and leading, for instance, after random unitarily evolving Zurek's initial
state configuration
to a completely mixed stationary state
with vanishing mutual information for all values of the fraction parameter.

In the random unitary evolution model pure dephasing
does not contribute to the structure of the attractor spaces emerging
from iterative application of pure decoherence (CNOT)
or of dissipative operations.
This means that the (dis-)appearance of quantum Darwinism in the asymptotic limit of
the random unitary model depends significantly on the interplay between the
dissipative and the CNOT-part of Eq. (\ref{Gl 2.4}).

The order of application of CNOT and Eq. (\ref{Gl 2.3})
in Eq. (\ref{Gl 2.4}) is irrelevant for the (dis-)appearance of quantum Darwinism in
the asymptotic limit of the random unitary model. Quantum Darwinism appears in the random unitary model only in case of
vanishing dissipation. There is no system-environment initial state within the dissipative-dephasing
random unitary model that leads to quantum Darwinism in case of non-vanishing dissipation.

It is an interesting problem for further research to explore also quantum Darwinism and its modification in even more general contexts, such as in the case of non-unitary randomly applied two qubit operations.

\paragraph{Acknowledgements}

~

The authors would like to thank G. Alber and J. Novotn$\textrm{\ensuremath{\acute{y}}}$ for stimulating discussions.

\paragraph{Author contribution statement}

~

The explicit analytic calculation of the dissipative-dephased attractor space was obtained by N. Balaneskovi$\textrm{\textrm{\ensuremath{\acute{\textrm{c}}}}}$ in the framework of his PhD-research. The numerical results of this paper (Fig. \ref{Fig.3} - Fig. \ref{Fig.6}) were obtained by  M. Mendler in the framework of his BSc-Thesis.
%
\twocolumn[{%

\appendix

\section{$S$-$E$-output states for the asymmetric dissipation in Zurek's qubit model \label{AAG}}

~~~In this appendix we derive generalized versions of states from Eq. (\ref{Gl 5.1}) discussed in the framework of Zurek's qubit model of quantum Darwinism with respect to asymmetric dissipation.

For $\left(\alpha_{1}\neq\alpha_{2}>0,\,\gamma=0\right)$ we
see that in Eq. (\ref{Gl 2.4}) only terms \[A=\cos\left(\frac{\alpha_{1}-\alpha_{2}}{2}\right)\left|0\right\rangle _{i}\left\langle 0\right|\otimes\left|0\right\rangle _{j}\left\langle 0\right|,\,
B=i\sin\left(\frac{\alpha_{1}+\alpha_{2}}{2}\right)\left|0\right\rangle _{i}\left\langle 1\right|\otimes\left|1\right\rangle _{j}\left\langle 0\right|,\,
C=i\sin\left(\frac{\alpha_{1}-\alpha_{2}}{2}\right)\left|1\right\rangle _{i}\left\langle 0\right|\otimes\left|1\right\rangle _{j}\left\langle 0\right|\]
and \[D=\cos\left(\frac{\alpha_{1}+\alpha_{2}}{2}\right)\left|1\right\rangle _{i}\left\langle 1\right|\otimes\left|0\right\rangle _{j}\left\langle 0\right|\]
(which vanishes for $\alpha_{1}=2\pi/3$ and $\alpha_{2}=\pi/3$ in
Eq. (\ref{Gl 5.1}) above) for $i\in\left\{ 1\right\} $ and $j\in\left\{ 1,\,...,\, n\right\} $
contribute when acting upon $\left|\Psi_{S}^{\textrm{in}}\right\rangle \otimes\left|0_{n}\right\rangle $
from Eq. (\ref{Gl 3.5}) within Zurek's qubit-model. In order to anticipate
those combinations of these four operators that act non-trivially
upon $\left|\Psi_{S}^{\textrm{in}}\right\rangle \otimes\left|0_{n}\right\rangle $
in Eq. (\ref{Gl 5.1}) we take into account only
allowed changes of the $S$-qubit states $\left\{ \left|0\right\rangle _{i},\,\left|1\right\rangle _{i}\right\} $
due to $\left(A+B+C+D\right)^{n}$ for $L=n\geq1$.
For instance, for $L=n=2$ the only allowed
sequences of the four operators acting in a non-trivial way upon the
$S$-states $\left|0\right\rangle _{i}$ and $\left|1\right\rangle _{i}$,
respectively, are $\left\{ A^{2},\, CA,\, BC,\, DC\right\} $
and $\left\{ AB,\, CB,\, BD,\, D^{2}\right\} $. These sequences generate the state
\begin{equation}
\begin{array}{l}
\left|\Psi_{SE_{L}}^{\textrm{out}}\right\rangle =a\left[\cos^{2}\left(\frac{\alpha_{1}-\alpha_{2}}{2}\right)\left|0\right\rangle _{i}+i\cos\left(\frac{\alpha_{1}-\alpha_{2}}{2}\right)\sin\left(\frac{\alpha_{1}-\alpha_{2}}{2}\right)\left|1\right\rangle _{i}\right]\left|0_{2}\right\rangle \\
+b\left[i\cos\left(\frac{\alpha_{1}-\alpha_{2}}{2}\right)\sin\left(\frac{\alpha_{1}+\alpha_{2}}{2}\right)\left|0\right\rangle _{i}+i^{2}\sin\left(\frac{\alpha_{1}+\alpha_{2}}{2}\right)\sin\left(\frac{\alpha_{1}-\alpha_{2}}{2}\right)\left|1\right\rangle _{i}\right]\left|10\right\rangle \\
+a\left[i^{2}\sin\left(\frac{\alpha_{1}+\alpha_{2}}{2}\right)\sin\left(\frac{\alpha_{1}-\alpha_{2}}{2}\right)\left|0\right\rangle _{i}+i\sin\left(\frac{\alpha_{1}-\alpha_{2}}{2}\right)\cos\left(\frac{\alpha_{1}+\alpha_{2}}{2}\right)\left|1\right\rangle _{i}\right]\left|01\right\rangle \\
+b\left[i\sin\left(\frac{\alpha_{1}+\alpha_{2}}{2}\right)\cos\left(\frac{\alpha_{1}+\alpha_{2}}{2}\right)\left|0\right\rangle _{i}+\cos^{2}\left(\frac{\alpha_{1}+\alpha_{2}}{2}\right)\left|1\right\rangle _{i}\right]\left|11\right\rangle 
\end{array}\label{Gl G2}
\end{equation}
as a general version of Eq. (\ref{Gl 5.1}), which induces $H\left(S\right)<H\left(S_{\textrm{class}}\right)\forall n\geq1$,
since for $L=n>2$ the allowed operator sequences lead to $\left|\Psi_{SE_{L}}^{\textrm{out}}\right\rangle $ from Eq. (\ref{Gl 5.1})
equivalent to Eq. (\ref{Gl G2}) for $\alpha_{1}=2\pi/3$ and $\alpha_{2}=\pi/3$. Certainly, for $\alpha_{1}=\alpha_{2}=0$
Eq. (\ref{Gl G2}) coincides with Eq. (\ref{Gl 3.5}). Accordingly, for the
operators \[A_{R}=A,\,B_{R}=i\sin\left(\frac{\alpha_{1}-\alpha_{2}}{2}\right)\left|0\right\rangle _{i}\left\langle 1\right|\otimes\left|0\right\rangle _{j}\left\langle 0\right|,\,
C_{R}=C,\,D_{R}=\cos\left(\frac{\alpha_{1}-\alpha_{2}}{2}\right)\left|1\right\rangle _{i}\left\langle 1\right|\otimes\left|1\right\rangle _{j}\left\langle 0\right|,\]
associated with a reversed order of CNOT and the dissipative-dephasing part in Eq. (\ref{Gl 2.4}), one can arrange allowed operator sequences in a similar way as done above, obtaining
for $L=n=2$ the state
\begin{equation}
\begin{array}{l}
\left|\Psi_{SE_{L}}^{\textrm{out}\left(R\right)}\right\rangle =\left[a\cos^{2}\left(\frac{\alpha_{1}-\alpha_{2}}{2}\right)+ib\cos\left(\frac{\alpha_{1}-\alpha_{2}}{2}\right)\sin\left(\frac{\alpha_{1}-\alpha_{2}}{2}\right)\right]\left|0\right\rangle _{i}\left|0_{2}\right\rangle \\
+\left[ib\cos\left(\frac{\alpha_{1}-\alpha_{2}}{2}\right)\sin\left(\frac{\alpha_{1}-\alpha_{2}}{2}\right)+i^{2}a\sin^{2}\left(\frac{\alpha_{1}-\alpha_{2}}{2}\right)\right]\left|0\right\rangle _{i}\left|10\right\rangle \\
+\left[i^{2}b\sin^{2}\left(\frac{\alpha_{1}-\alpha_{2}}{2}\right)+ia\sin\left(\frac{\alpha_{1}-\alpha_{2}}{2}\right)\cos\left(\frac{\alpha_{1}-\alpha_{2}}{2}\right)\right]\left|1\right\rangle _{i}\left|01\right\rangle \\
+\left[ia\sin\left(\frac{\alpha_{1}-\alpha_{2}}{2}\right)\cos\left(\frac{\alpha_{1}-\alpha_{2}}{2}\right)+b\cos^{2}\left(\frac{\alpha_{1}-\alpha_{2}}{2}\right)\right]\left|1\right\rangle _{i}\left|11\right\rangle 
\end{array}\label{Gl G3}
\end{equation}
as a generalized version of Eq. (\ref{Gl 5.1}), with $H\left(S\right)<H\left(S_{\textrm{class}}\right)\forall n\geq1$.
Again, for $\alpha_{1}=\alpha_{2}=0$ Eq. (\ref{Gl G3}) coincides with
Eq. (\ref{Gl 3.5}).

}]



\end{document}